\documentclass[preprint2]{emulateapj}

\shorttitle{Astrometry}
\shortauthors{Guyon et al.}
\usepackage{graphicx}
\usepackage[mathscr]{eucal}
\begin{document}

\title{HIGH PRECISION ASTROMETRY WITH A DIFFRACTIVE PUPIL TELESCOPE}

\author{Olivier Guyon}
\affil{Steward Observatory, University of Arizona, Tucson, AZ 85721, USA}
\affil{National Astronomical Observatory of Japan, Subaru Telescope, Hilo, HI 96720, USA}
\email{guyon@naoj.org}
\author{Eduardo A. Bendek, Thomas D. Milster}
\affil{College of Optical Sciences, University of Arizona, Tucson, AZ 85721, USA}
\author{Josh A. Eisner, Roger Angel, Neville J. Woolf}
\affil{Steward Observatory, University of Arizona, Tucson, AZ 85721, USA}
\author{S. Mark Ammons}
\affil{Lawrence Livermore National Laboratory, Physics Division L-210,  7000 East Ave Livermore CA 94550, USA}
\author{Michael Shao, Stuart Shaklan, Marie Levine, Bijan Nemati}
\affil{Jet Propulsion Laboratory, 4800 Oak Grove Drive, Pasadena, CA 91109, USA}
\author{Joe Pitman}
\affil{Exploration Sciences, PO Box 24, Pine, CO 80470, USA}
\author{Robert A. Woodruff}
\affil{2081 Evergreen Avenue, Boulder, CO 80304, USA}
\author{Ruslan Belikov}
\affil{NASA Ames Research Center, Moffett Field, CA 94035, USA}

\begin{abstract}
Astrometric detection and mass determination of Earth-mass exoplanets requires sub-$\mu$as accuracy, which is theoretically possible with an imaging space telescope using field stars as an astrometric reference. The measurement must however overcome astrometric distortions which are much larger than the photon noise limit. To address this issue, we propose to generate faint stellar diffraction spikes using a teo-dimensional grid of regularly spaced small dark spots added to the surface of the primary mirror (PM). Accurate astrometric motion of the host star is obtained by comparing the position of the spikes to the background field stars. The spikes do not contribute to scattered light in the central part of the field and therefore allow unperturbed coronagraphic observation of the star's immediate surrounding. Because the diffraction spikes are created on the PM and imaged on the same focal plane detector as the background stars, astrometric distortions affect equally the diffraction spikes and the background stars, and are therefore calibrated. 
We describe the technique, detail how the data collected by the wide-field camera are used to derive astrometric motion, and identify the main sources of astrometric error using numerical simulations and analytical derivations. We find that the 1.4 m diameter telescope, 0.3 deg$^2$ field we adopt as a baseline design achieves 0.2 $\mu$as single measurement astrometric accuracy. The diffractive pupil concept thus enables sub-$\mu$as astrometry without relying on the accurate pointing, external metrology or high stability hardware required with previously proposed high precision astrometry concepts.  
\end{abstract}

\section{Introduction}
\label{sec:intro}

\subsection{Background}
Detection and characterization of potentially habitable Earth-mass exoplanets is one of the leading astronomical challenges of our age. Thanks to indirect detection techniques, such as radial velocity, transit photometry, and microlensing, the number of known exoplanets is rapidly growing, providing valuable information on the frequency and diversity of exoplanets. Two approaches --- direct planet imaging and host star astrometry --- have the potential for revolutionary discoveries by obtaining a complete census of exoplanets around nearby stars and characterizing them:
\begin{enumerate}
\item{Direct imaging of exoplanets with future space telescopes will reveal their atmospheric composition and possibly identify signs of biological activity \citep{2009arXiv0911.3200L}.}
\item{Astrometry, by providing a precise measurement of the host star motion on the sky will yield the planet mass and orbit \citep{2005ASPC..338...37U,2009astro2010S.271S}.}
\end{enumerate}
While either technique is suitable to identify nearby planets, both are required for unambiguous characterization of potentially habitable worlds \citep{2010ApJ...720..357S}. It has been so far assumed that coronagraphic imaging/spectroscopic measurement and mass determination require two separate missions. An approach is proposed here that combines the two techniques using a single space telescope in which light is simultaneously fed to a narrow field coronagraph for exoplanet imaging and spectroscopy, and a wide-field astrometric camera imaging a wide annulus around the central field for mass measurement with astrometry. 

\begin{figure*}
\includegraphics[scale=0.45]{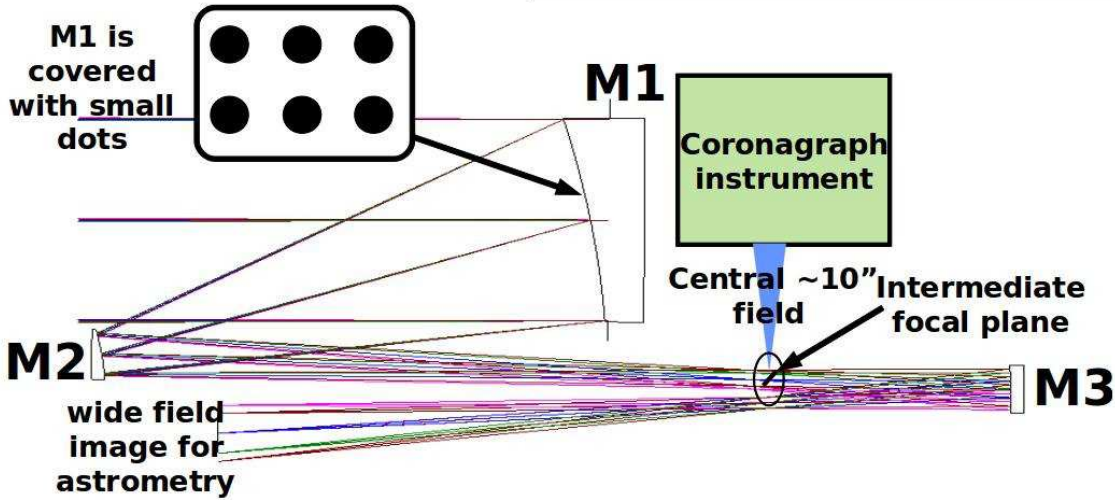} 
\caption{\label{fig:optprinciple} Conceptual optical design for a space telescope combining deep wide-field imaging, exoplanet coronagraphy, and astrometry measurements. The wide-field three-mirror telescope produces a well-corrected diffraction-limited image on a wide-field focal plane array (bottom left). The central part of the field, containing both the central star and its exoplanet(s), is fed to a coronagraph instrument thanks to a small pickoff mirror in the telescope's intermediate focus. Small dots regularly spaced on the primary mirror surface produce a diffraction pattern in the wide-field image which allows accurate astrometric referencing of the central star to background field stars.}
\end{figure*}

\subsection{Brief overview of the technique}
Astrometric measurement from wide-field images is fundamentally limited, in a perfect system, by photon noise and sampling effects, which are quantified in the Appendix and taken into account in this paper for numerical performance estimates. These fundamental limits are however not the focus of this paper, which is aimed at providing a solution to the three main practical challenges to performing precision absolute astrometry of a bright star from a wide-field image when using numerous faint field stars as the astrometric reference.
\begin{itemize}
\item{\textit{Dynamical range}. There is a large brightness difference between the central target star and the surrounding field stars, making it difficult for a detector to properly image both.}
\item{\textit{Distortions}. Slight deformations of the optical surfaces, or the detector focal plane array, introduce astrometric errors. These errors tend to grow larger in amplitude as the FOV is increased.}
\item{\textit{Detector defects}. The geometry and response of pixels is not known to sufficient accuracy to allow high-precision astrometry from a single wide-field image.}
\end{itemize}

The proposed diffractive pupil astrometry optical principle, detailed in Section \ref{sec:principle}, solves the first two challenges by creating in the wide-field focal plane image diffraction spikes. These spikes are of comparable surface brightness as the field stars (solves the dynamical range challenge), and experience the same distortions as the field stars used as the astrometric reference (solves the distortion challenge).

Detector defects are not calibrated by the diffraction spikes, and require averaging of many measurements to be reduced to the desired sub-$\mu$as level. Our proposed concept uses a large number (typically hundreds to thousands) of faint ($m_V \approx 14$ and fainter) field stars for astrometric referencing, and therefore differs from interferometric approaches \citep{2005ASPC..338...37U} performing pairwise astrometric measurements using bright stars. The large number of stars used for astrometric referencing averages down by one to two orders of magnitudes high-order astrometric errors. As described in Section \ref{sec:telescoperoll}, further averaging is required for sub-$\mu$as precision, and is achieved by continuously rolling the telescope during observations. 

In Section \ref{sec:dataacquproc}, we describe how the astrometric data are acquired and how the measured positions of the field stars are combined toward the final astrometric measurement. Section \ref{sec:simerrbudg} provides an estimate of the astrometric precision for a medium-sized 1.4 m space telescope with a 0.3 deg$^2$ (0.6 deg diameter) FOV.

\section{Optical Principle}
\label{sec:principle}

\subsection{Optical Design Overview}

As shown in Figure \ref{fig:optprinciple}, the proposed technique uses a conventional wide-field diffraction-limited imaging telescope. The baseline telescope design adopted in this paper uses three mirrors (all conics) to optimize image quality over a wide FOV, and is an off-axis system to allow compatibility with all high performance coronagraph concepts. The central portion of the field is used for coronagraphy and reflected into a coronagraph instrument by a small pickoff mirror in the intermediate focal plane of the system. The rest of the field is imaged by an unfiltered (sensitive from 0.4 $\mu$m to 0.8 $\mu$m) wide-field diffraction-limited CCD camera at F/40, Nyquist sampled at 0.6 $\mu$m. The wide-field off-axis 1.4 m diameter telescope design shown in Figure \ref{fig:optprinciple} produces a 0.5 deg $\times$ 0.5 deg diffraction-limited wide-field image for astrometric measurement and feeds a coronagraph instrument with a narrow FOV extracted in the intermediate focus. The total on-axis point-spread function (PSF) diameter at the intermediate focus is 6 arcsec: the pick-off mirror must be at least 6 arcsec diameter to avoid vigneting, even for a very small coronagraph FOV. The telescope design was performed with the CodeV software, and then coded in a custom C program developed by our team to allow high-precision astrometric computations.

While the telescope diameter and design adopted are chosen to be cost-realistic and inspired from the Pupil mapping Exoplanet Coronagraphic Observatory mission concept study \citep{2010SPIE.7731E..68G}, the technique is applicable to other telescope sizes and optical design.

We show in Section \ref{ssec:dots} that, by placing non-reflective dots on the primary mirror (PM), diffraction spikes are created in the wide-field astrometric image to provide a suitable reference (linked to the central star) against which the position of the field stars is accurately measured. We show in Section \ref{ssec:fielddistortions} that all astrometric distortions (due, for example, to deformations of the secondary and tertiary mirrors, or deformations of the focal plane array) are by design common to the spikes and the background stars, and a differential astrometric measurement between the diffraction spikes and the field stars is therefore largely immune to large-scale astrometric distortions. Our proposed concept therefore does not require the picometer level stability (or metrology calibration) on the optics over years which would otherwise be essential for wide-field astrometric imaging \citep{2005ASPC..338...37U,2009SPIE.7439E..30M}.

\subsection{Dots on Primary mirrors, Spikes in the Wide-Field Astrometric Camera}

\label{ssec:dots}
\begin{figure*}
\includegraphics[scale=0.53]{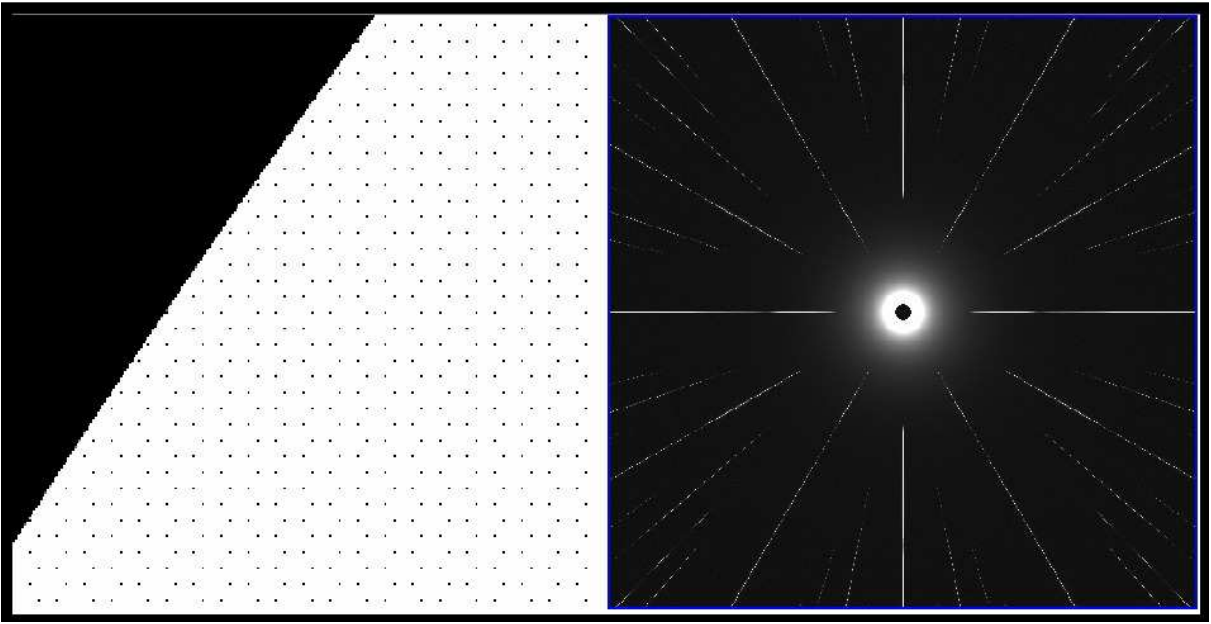} 
\caption{\label{fig:principle} 
Dots on the telescope primary mirror (left) and corresponding on-axis PSF in the wide-field astrometric camera (right). The primary mirror area shown is 3\% of the pupil diameter across, and is located at the edge of the pupil. The spacing between the diffraction spikes, their extent in the focal plane, and their overall luminosity can be chosen by appropriate design of the dot pattern on the pupil.}
\end{figure*}

As shown in Figure \ref{fig:principle} (left), a regular grid of dark (non-reflective) spots is physically etched/engraved on the front surface of the PM. The dots act as a two-dimensional (2D) diffraction grating: the monochromatic system PSF is an Airy pattern surrounded by a widely spaced grid of fainter Airy patterns. In polychromatic light, the secondary Airy patterns are radially dispersed, producing the long diffraction spikes visible in Figure \ref{fig:principle} (right). This PSF appears at the focal plane for each field object displaced so it is centered where its star is imaged, respectively, modulated in brightness by the source magnitude. The spikes from the field stars are therefore very dim, while the spikes from the host star are much brighter. Light in the central part of the field is directed to the coronagraph, therefore suppressing its bright central Airy pattern while passing its polychromatic diffraction pattern (aka spikes) to the focal plane array used for astrometry. When the telescope is pointed at a bright star, these spikes will be superimposed on a background of numerous faint field stars used as the astrometric reference. Precise measurement of the position of the bright central star against this background reference is possible by simultaneously imaging on a diffraction-limited wide-field camera both the spikes and the background of faint reference stars. Our approach uses only the images acquired with the wide-field camera to perform the relative astrometric measurement between the central star and a reference consisting of a large number of background faint stars: the central field which can be directed to the coronagraph instrument does not contribute to the astrometric measurement. The system is thus immune to flexure or non-common path errors between the two optical paths, and there is no need to reference the two paths.

The spikes provide an adequately bright signal for the host star light, solving the contrast problem between the bright central star and the fainter background field stars used as the astrometric reference. By distributing a few percent of the host star's light over a large number of pixels, the spikes provide a feature that can be imaged without saturation on the same detector as the background stars. A similar magnitude compensation scheme using a grating in front of the telescope has previously been used over small angles for ground-based astrometry of binary stars \citep{1946AJ.....52....1S}, and more recently with a grating in a relay pupil for coronagraphic imaging and astrometry of faint companions with adaptive optics \citep{2006ApJ...647..620S,2006ApJ...647..612M,2010ApJ...709..733Z}. We note that the schemes proposed by \cite{2006ApJ...647..620S} and \cite{2006ApJ...647..612M} are fundamentally different in goals from our proposed diffractive pupil concept, as (1) they are aimed at performing a relative astrometric measurement between a star and its faint companion in a coronagraph, while our concept is aimed at measuring the absolute position of a star against a reference consisting of other stars in a wide-field image, (2) they do not allow calibration of wide-field distortions which are introduced by large optical elements (main telescope optics), as this requires the first optical element to be the diffractive element.

\subsection{Immunity to Field Distortions}
\label{ssec:fielddistortions}

In a conventional wide-field telescope, a high-precision astrometric measurement is not possible, as small unknown errors in the shape of the optics create astrometric distortions: for different positions on the sky, the beam footprint on optical elements which are not conjugated to the pupil is different, producing tip-tilt anisoplanatism (aka astrometric distortions). Physical distortions of the focal plane array also contribute to astrometric distortions.


\begin{figure*}
\includegraphics[scale=0.64]{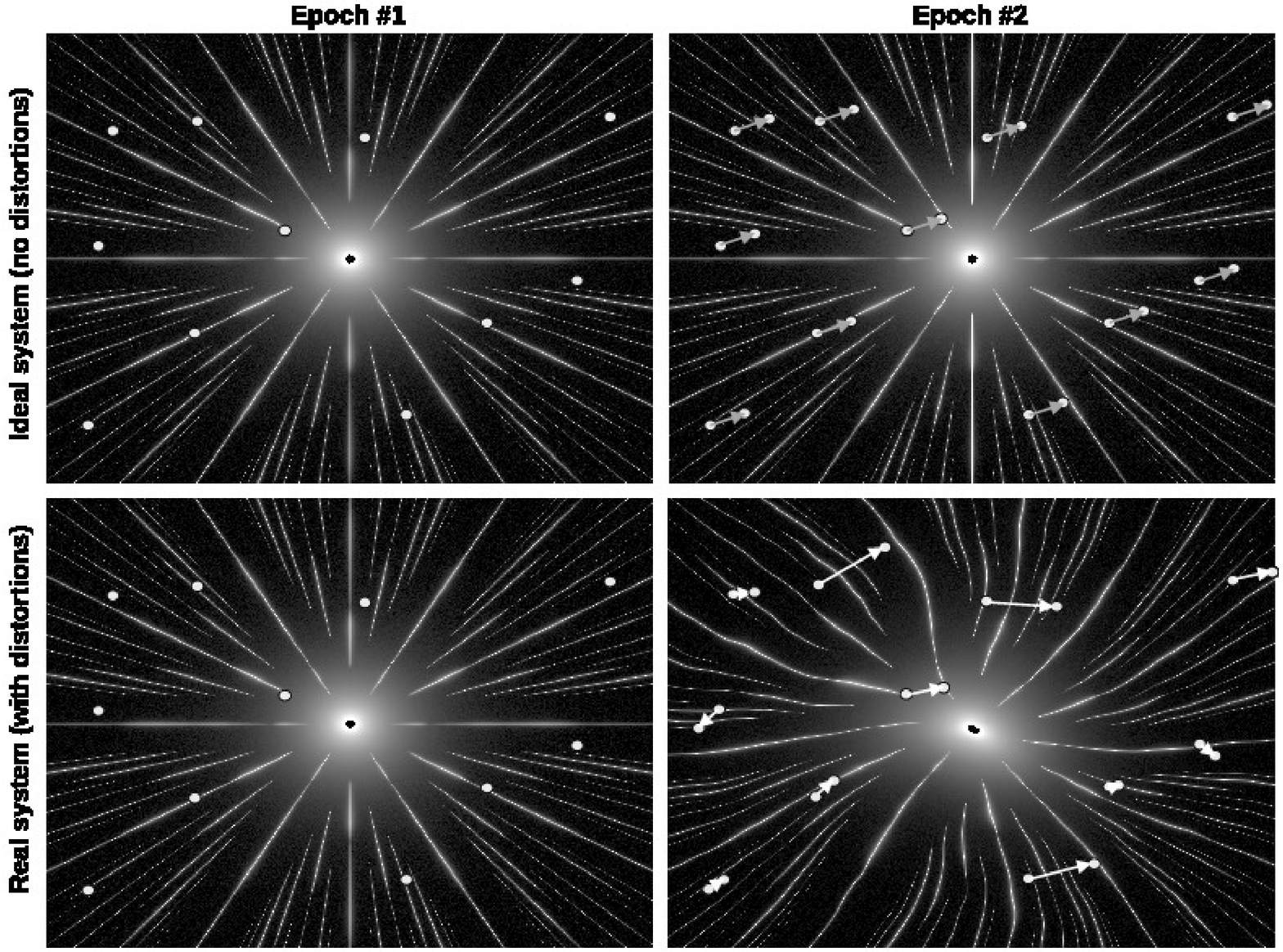} 
\caption{\label{fig:WFimages}
Calibration of astrometric distortions with diffraction spikes in the focal plane. In a perfectly static optical imaging system (top row), the astrometric motion of the central star is recorded as a change in the position of the background stars relative to the central star. This motion is shown as parallel arrows in the top right image. In this case, the role of the diffraction spikes is simply to provide an adequately bright image feature referenced to the central star (which is missing from the images since its light has been blocked and/or sent to a coronagraph instrument). The bottom rows show how the same astrometric measurement is affected by a change in the imaging system astrometric distortion. The arrows show the measured motion of the background stars relative to the central star. This motion is, for each background star, equal to the sum of the central stars' astrometric motion (to be measured) and the much larger change in astrometric distortions between the two observation epochs. The diffraction spikes are also affected by the same distortion change, which changes their shape. By comparing the images of the diffraction spikes between the two epochs, the change in astrometric distortion can be calibrated and compensated for. Each image is a simulated 0.2 $\times$ 0.2 deg field imaged by a 1.4 m diameter telescope, and the spikes image shown in the lower right was obtained by amplifying by 1e6 the astrometric distortion computed by raytracing through the optical system shown in Figure \ref{fig:optprinciple}, with realistic surface errors on the telescope optics. The amplitude of the astrometric motion used in this figure (length of the arrows in the top right figure) has also been greatly amplified for clarity.}
\end{figure*}

The diffraction spikes created by the diffractive pupil encode all instrumental distortions since the reference pattern (diffraction spikes) is introduced directly on the PM of the telescope. The dots on the PM act as a diffraction grating creating secondary beams which emerge from the primary mirror with slightly different angles and travel through the optical system up to the focal plane. Light from an off-axis star and light from a nearby diffraction spike go through nearly the same path in the optical system (telescope + instrument) and share the same astrometric distortion. The anisoplanatism problem is therefore eliminated in the differential spike/background star astrometric measurement, as illustrated in Figure \ref{fig:WFimages}. The proposed scheme also calibrates focal plane array distortions, as they will affect equally the background stars and the diffraction spikes. We note that wavefront errors on the PM and telescope pointing errors do not directly produce an astrometric error as they are common to both the diffracted beams and the beam from the astrometric reference stars.


For the spikes to encode exactly the same astrometric distortions as the background field stars, the telescope design must satisfy the following criteria.
\begin{itemize}
\item{The dots must cover uniformly the PM, otherwise, changes in PM shape can create a differential motion between the spikes and the background stars. For example, if the dots cover only a zone of the PM, the spikes will move with the average wavefront slope over the area of the PM covered with dots, while background stars will move with the overall wavefront slope over the whole PM.}
\item{The PM must be the aperture stop for the system, so that aberrations introduced by the PM have no field dependence, and therefore produce no astrometric distortions (the diffraction spikes can only calibrate distortions introduced by optics after the plane in which the dots are placed).}
\item{There must not be any chromatic optics between the PM and the wide-field camera detector. Refractive optics have some chromaticity, and the spikes are chromatically elongated (a background star and a spike near it therefore have very different colors, and could see different distortions in a system with refractive optics). With refractive elements, the color of background stars would need to be known in order to identify which parts of the spikes should be used for high precision astrometry.}
\end{itemize}

The concept discussed in this paper fulfills these three requirements. For implementations that do not fulfill these requirements, additional error term(s) would need to be quantified, and the estimated astrometric precision obtained in this paper would therefore not be applicable.

\begin{figure}
\includegraphics[scale=0.47]{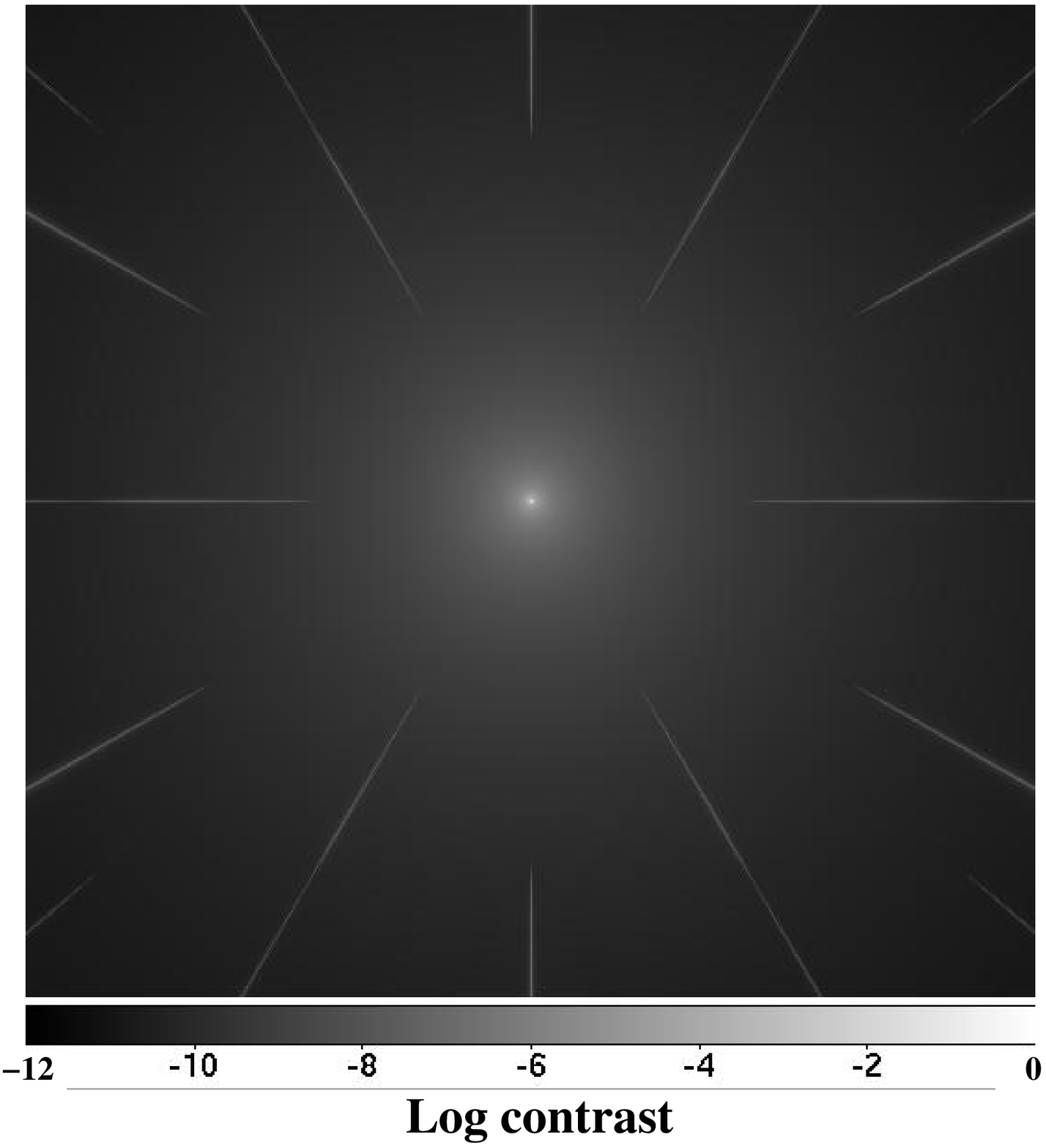} 
\caption{\label{fig:psfcenter}
Central part of the PSF (log scale, 3 arcmin $\times$ 3 arcmin). The simulated polychromatic PSF is shown here with no coronagraph pickoff mirror. In this configuration, the diffraction spikes do not affect the central few arcsecond around the optical axis, and have no effect on coronagraphic performance. The halo around the central star is due to the PSF's Airy rings.}
\end{figure}

\subsection{Simultaneous operation with a coronagraph}

Since the PM mask is by design a regular grid containing no low-order aberrations, it does not impact high contrast coronagraph observation performed by a separate narrow field instrument (see Figure \ref{fig:psfcenter}), other than a small loss in throughput: as seen by the coronagraph, the pupil is uniformly gray, with a few percent of the light missing. The effect of the dots is therefore equivalent to a uniform loss in reflectivity in the coating. Small errors in the manufacturing of the dot pattern will however introduce low-order amplitude modulations in the pupil, and must be kept sufficiently small to be adequately removed by the coronagraph's wavefront control system.

Another potential issue for the coronagraph is the presence of diffraction spikes from other stars in the FOV of the wide angle camera. Thankfully, the spikes are faint (approximately 1e-8 of the surface brightness of the peak PSF for the baseline concept adopted in this paper), and occupy a small area of the focal plane. In the pessimistic case where a similarly bright star is within the field, the average surface brightness in the coronagraphic FOV is equal to the fractional area covered by the dots divided by the camera FOV (area over which the spikes extend) in unit of $(\lambda/D)^2$. With the 1.4 m telescope example considered in this paper, 0.29 deg$^2$ ($=$ 0.6 deg diameter) FOV, and 1\% dot coverage, the average surface brightness in the coronagraphic FOV is at the 2e-11 contrast level, significantly below the expected zodiacal and exozodiacal contrast levels. The effects of spikes can further be reduced by removing from the data set the small fraction of the observations when the spikes are known to move across the central field as the telescope is rolling.

Performing astrometry and coronagraphy with the same telescope is an efficient combination, as both techniques require a stable telescope delivering stable PSFs. A coronagraph mission would observe a small number of bright targets with long exposure times and several visits; this observation mode is also suitable for the astrometric measurement.

\section{Continuous Telescope Roll}
\label{sec:telescoperoll}

The diffraction spikes encode low-order astrometric distortions, defined here as distortions that vary on a scale equal or larger than the separation between spikes. High-order astrometric distortions, including detector effects, cannot be calibrated using the diffractive spikes. 
These errors must be further reduced by averaging many non-correlated measurements in order to achieve sub-$\mu$as astrometric accuracy without enforcing challenging requirements on the focal plane detector. While the large number of background stars used as the astrometric reference provides some averaging, it is not sufficient to achieve sub-$\mu$as accuracy, and we propose to continuously roll the telescope around the optical axis during observations to achieve this goal.

\subsection{Roll versus Fixed pixel position approaches}

Unknown detector imperfections have a large effect on astrometric accuracy and must be taken into consideration in the data acquisition. There are two possible approaches to mitigate this problem.
\begin{itemize}
\item{Keeping the star(s) on nearly the same pixel position between measurements to perform a differential measurement which is insensitive to static detector imperfections, or,}
\item{Averaging down detector defects by performing a large number of measurements over different pixels.}
\end{itemize}

The first approach requires a very specific design  \citep{2005ASPC..338...37U}, and would be required to change the detector geometry to follow field stars in an astrometric imaging system. It is otherwise not possible to keep background stars on exactly the same pixels of the detector during the measurement timescale due to the central star's proper motion (which can drag the background stars at approximately 1 arcsec year$^{-1}$), its parallax, and the aberration of light. Moreover, using the same limited number of pixels for the astrometric measurement requires very good long term stability of their response.

\begin{figure}
\includegraphics[scale=0.4]{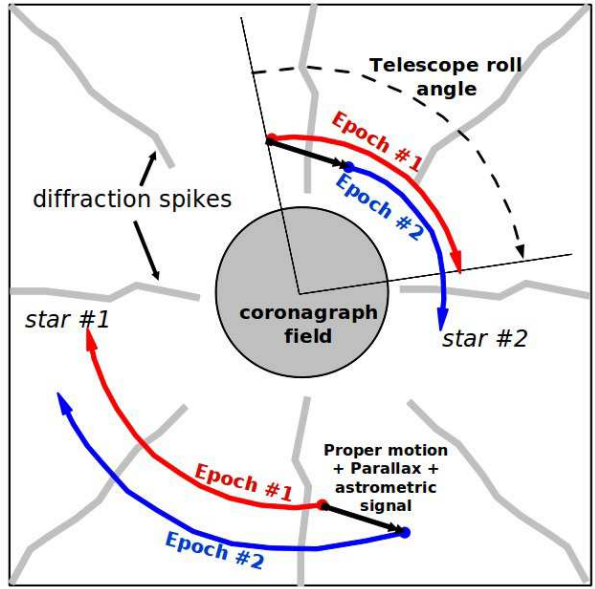} 
\caption{\label{fig:rollgeom} 
Telescope roll geometry. Position of two background stars on the camera focal plane during telescope roll. Observations at two epochs (colored red and blue) are shown for the two stars. The diffraction spikes position on the detector is fixed, while background stars are moving.
}
\end{figure}

If background stars cannot practically be kept on the same pixels, each measurement of the motion of a background star between two epochs compares PSFs falling on different pixels with different unknown characteristics (pixel sensitivity, size, shape, etc.). Many statistically independent measurements are required to average this error term, as the desired sub-$\mu$as accuracy corresponds to approximately $10^{-5}$ pixel: with a 1/100th pixel single star single measurement centroiding accuracy, a $\approx 10^{3}$ averaging factor (obtained with $10^6$ uncorrelated measurements) is required to reach sub-$\mu$as accuracy. By averaging measurements from a few hundred stars, an averaging factor $\approx 10$ is obtained, and the final measurement is still $\approx 100$ times short of the goal. We therefore achieve the averaging by both combining the position measurements of a large number of background field stars and rolling the telescope along the line of sight to move the background stars' PSFs over a large number of pixels. With a $\approx$ 1 rad roll, stars will move over $\approx 10^4$ pixels, providing the required 100$\times$ averaging assuming uncorrelated measurements. As described in Section \ref{ssec:rollanticorrelation}, the averaging gain is significantly better than this thanks to a strong roll anticorrelation of the detector errors. The proposed roll geometry is shown in Figure \ref{fig:rollgeom}.

The telescope roll is very efficient at reducing the contribution of detector errors to the final astrometric measurement. During a single observation (typically a day), the telescope is slowly rolled around the line of sight to move the background PSFs on the focal plane. On a large format detector, a roll angle of a few tens of degrees is large enough to move the PSF by several thousand pixels, and a 1 rad roll will produce about 10,000 PSF centroid measurements per star.

\begin{figure}
\includegraphics[scale=0.23]{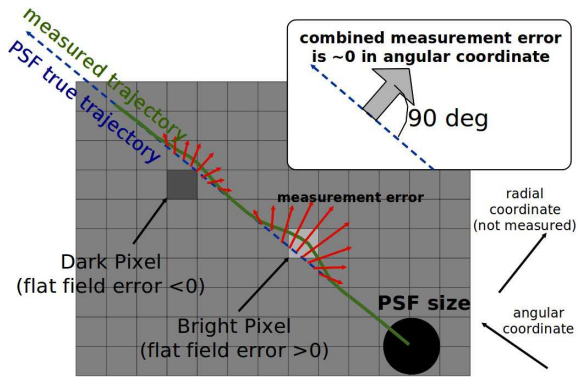} 
\caption{\label{fig:rollanticorr} 
Effect of flat-field errors between pixels on the astrometric measurement of a PSF moving in a straight line. The true PSF trajectory (dashed line) differs from the measured PSF position (continuous curve) due to the presence of bright and dark pixels. When compensated for the telescope roll, the combined measurement error is perpendicular to the PSF motion: there is no error along the direction of the PSF motion. 
}
\end{figure}

\subsection{Roll Anticorrelation of astrometric errors}
\label{ssec:rollanticorrelation}

Unknown flat-field errors produce astrometric measurement errors: a pixel which is more sensitive than assumed will "attract" the measured position while pixels less sensitive than assumed will "push away" the measured position. Figure \ref{fig:rollanticorr} shows how this measurement error (red arrows) evolves as the PSF drifts across the detector during telescope roll. The error can be decomposed into a radial component (along the direction to the central star) and an azimuthal component. When integrated over time as the telescope rolls, the residual error is almost entirely radial, as the azimuthal error when the PSF approaches the defective pixel is compensated by the opposite error when the PSF drifts away from the defective pixel. The telescope roll can therefore almost entirely remove flat field induced errors along the axis where the astrometric measurement is performed. We note that, as detailed in Section \ref{sec:dataacquproc}, the position measurement in the direction along the spikes is not used, as the spikes elongation is not suitable for distortion measurement along this axis, and the roll anticorrelation described in this section does not apply.

The discussion above assumes sensitivity differences between pixels, but is also valid for other differences between pixels, such as a pixel with a peculiar color preference, or a pixel with a dead corner. The key advantage of the roll is that it transforms detector defect induced astrometric errors into a time-variable error which is strongly anticorrelated on both sides of the defect. These errors therefore disappear in the averaged measurement (this is much better than a decorrelated error which slowly decreases as $\sqrt{N}$).

\section{Data Acquisition and Processing}
\label{sec:dataacquproc}

\begin{figure*}
\includegraphics[scale=0.32]{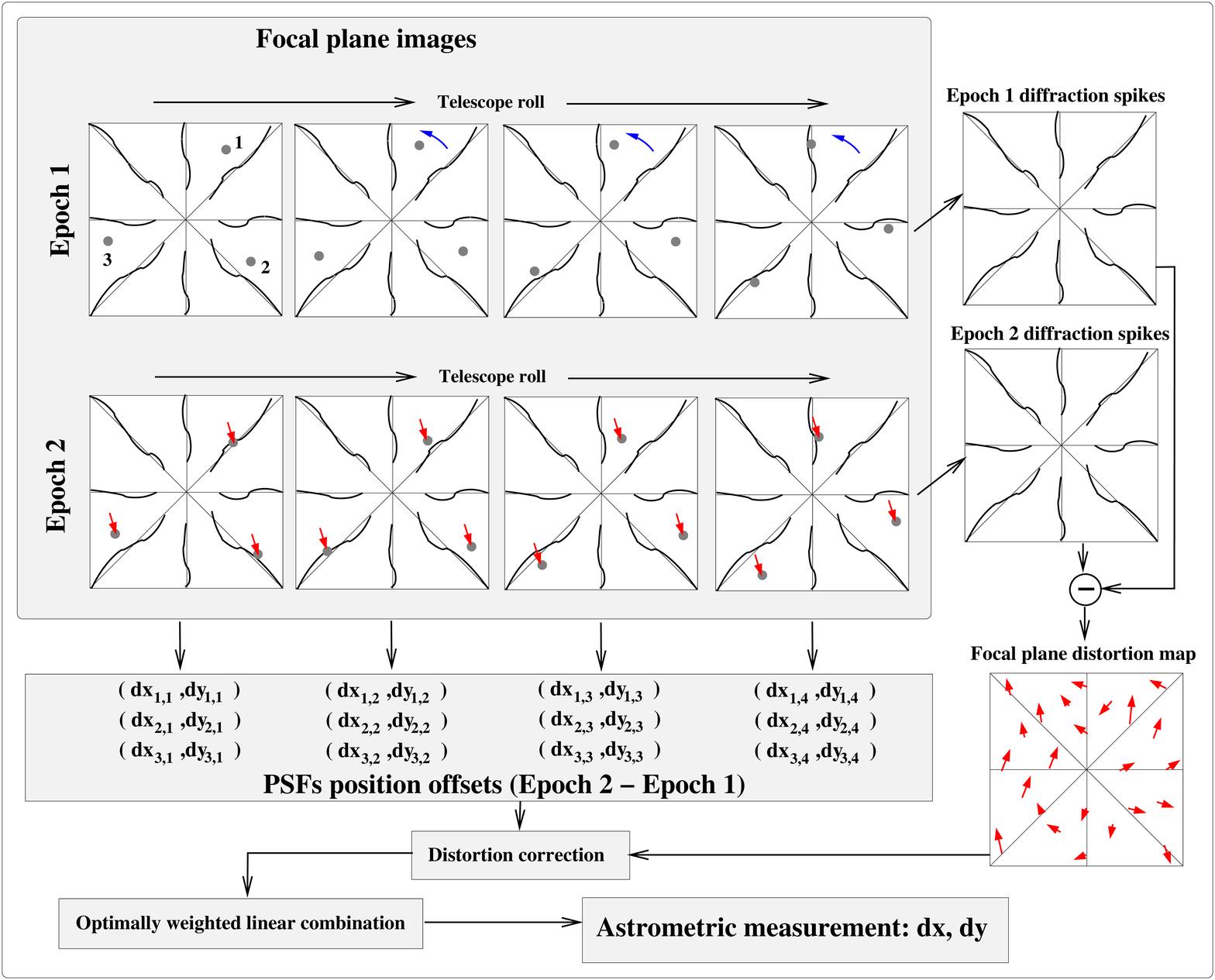} 
\caption{\label{fig:dataacqu} 
Data acquisition and processing overview.}
\end{figure*}

\subsection{Data acquired}
The data acquired consist, for each observation epoch, of a series of exposures acquired while the telescope is slowly rolling around the line of sight. The individual exposures are kept sufficiently short to avoid smearing of the PSFs at the edge of FOV by more than approximately 1 pixel, but sufficiently long to ensure photon noise limited sensitivity. A full sequence of exposures spans several hours to days, while observing epochs are separated by weeks to months. The top left part of Figure \ref{fig:dataacqu} illustrates the data acquired for two epochs. The top row shows a sequence of four images acquired at epoch 1 (in an actual observation, the number of individual images would be much greater). The telescope is rolled between each image. While the diffraction spikes are fixed on the detector, the background stars (gray spots labeled 1, 2, and 3 in the top left image of the figure) rotate around the line of sight due to telescope roll. The same sequence of measurements is repeated at epoch 2 (second row). 

The astrometric measurement is performed differentially, between sets of images acquired at different epochs, and must detect a relative motion (translation) between the central star and the field stars. The precise position of the central star is not directly measured; its diffraction spikes are used instead. In the absence of astrometric distortions, the astrometric signal would be a pure field-invariant translation between the two epochs, which would be entirely derived from the measurement of the positions of the field stars on the detector, as described in Section \ref{ssec:measurement1}. As described in Section \ref{ssec:measurement2}, astrometric distortions are measured and corrected for by comparing images of the diffraction spikes between the two epochs.

\subsection{Measurement of PSFs position offsets between epochs}

\label{ssec:measurement1}

The displacement of the field stars between the two epochs is measured and shown as small red arrows in the epoch 2 images. The position of each field star on the detector is measured for each exposure by fitting a model of the PSF on a square grid detector. An optimal matched filter taking into account photon noise from both the stellar PSFs and the background (zodiacal light) is assumed for simulations and derivations in this paper, since it yields the best measurement accuracy. The effect of uncalibrated detector defects will be discussed in Section \ref{sec:simerrbudg}. A simpler photocenter algorithm may be used instead, but is expected to yield slightly reduced accuracy. We assume in this paper that the position measurement for the background stars is unaffected by the spikes; this approximation is not valid when the star image falls on a spike, which for any given background star occurs for a small fraction of the total observing time.

For the same telescope roll angle (same column in Figure \ref{fig:dataacqu}), the field stars are located almost (but not exactly) on the same position on the detector at the two epochs. This configuration is largely immune to uncalibrated/unknown PSF shape, as the measurement is differential and the PSF is not expected to change sufficiently over a sub-arcsec field to introduce an astrometric error. Maintaining nearly the same PSF positions between epochs also greatly reduces sensitivity to static astrometric distortions, which cannot be removed from the data (the diffraction spikes are used to measure changes in astrometric distortions, but do not measure absolute distortions). The position offset ($dx_{i,j}$,$dy_{i,j}$), the difference between field star position at two epochs on the focal plane array, is measured for each field star $i$ in the field for each roll angle position $j$ (for which the roll angle is denoted as $\theta_j$).

In the absence of measurement noises, the measured offsets are

\begin{equation}
dx_{i,j} = -dX \cos(\theta_j) - dY \sin(\theta_j)
\end{equation}
\begin{equation}
dy_{i,j} = -dX \sin(\theta_j) + dY \cos(\theta_j)
\end{equation}

where ($dX$,$dY$) is the astrometric motion of the central star between the two epochs and $\theta_j$ is the angle between the sky coordinate system ($X$,$Y$) and the camera coordinate system ($x$,$y$). 

\subsection{Calibrating astrometric distortions using diffraction spikes}
\label{ssec:measurement2}

Our proposed calibration of astrometric distortions relies on comparison between diffraction spikes images acquired at different epochs, and is therefore not sensitive to static astrometric distortions. Only changes in the astrometric distortions between epochs are measured. In the absence of such changes, the diffraction spikes would be kept at exactly the same position on the detector for all observations. Time-variable field distortions are measured by small changes in the location of the spikes, and are therefore only sampled along the spikes. 

The distortions can only be measured to high accuracy in the direction perpendicular to the spikes due to their strongly elongated shape. The radial component of the spikes motion cannot be measured to a good accuracy in the presence of noise. We also note that the highly beneficial anticorrelation effect described in Section \ref{ssec:rollanticorrelation} occurs only on the angular coordinate.
{\bf For each exposure, only the angular (perpendicular to the spikes) component of the background star's positions therefore contributes to the final astrometric measurement. Each individual astrometric measurement (single telescope roll angle, single background star) is one dimensional (1D), and the final 2D astrometric measurement is obtained by combining many 1D measurements}. 

A 2D interpolation is used to build a continuous 2D map of the distortion change (shown in the lower right of Figure \ref{fig:dataacqu}) between the observation epochs from the measurement of azimuthal distortion along the spikes. This azimuthal distortion map is then removed from the measurements. The final 2D astrometric measurement is obtained by combining all 1D azimuthal measurements with appropriate weighting coefficients (fainter stars where the photon noise contribution is large are given a smaller weight).

\section{Astrometric measurement: Simulations and error budget}
\label{sec:simerrbudg}

\subsection{Approach to performance estimation}
This section is aimed at providing an estimate of the astrometric precision of the proposed concept for a baseline design, for which a detailed numerical simulation is performed to quantify the contribution of several of key error terms.

A good understanding of how the astrometric error is affected by the brightness of the field stars is essential, as (1) the final astrometric measurement is obtained by combining measurements from background stars covering a large range of brightness and (2) the instrument design and data acquisition may need to be optimized for a relatively narrow range of background star brightness due to technical limitations (detector dynamical range). Our approach is thus to quantify the astrometric information offered by each field star (which is a function of field star brightness) and then to optimally combine the information from all field stars into a single 2D astrometric measurement.

\begin{figure}
\includegraphics[scale=0.43,angle=-90]{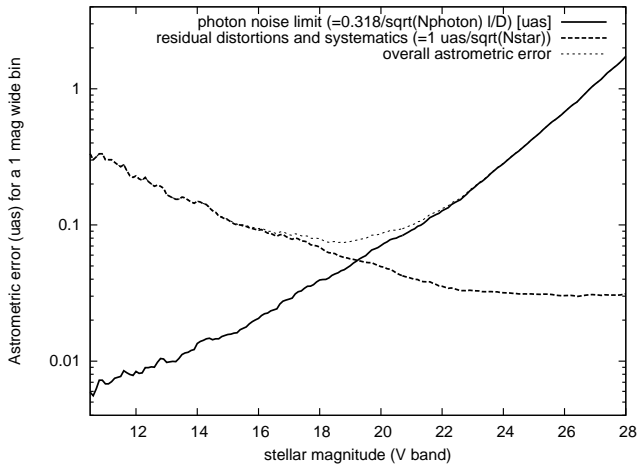} 
\caption{\label{fig:noisefact} Theoretical single axis astrometric measurement accuracy of a 1.4 m diameter telescope imaging a 1 deg$^2$ field (0.6 deg diameter) near the galactic pole with a 1 day observing time. The vertical axis shows the combined 1 $\sigma$ astrometric accuracy if all stars in a 1 mag wide brightness bin are used toward the measurement. Two noises are considered in this simple model: photon noise (assuming a perfect detector, no zodiacal light) and distortion noise which is set to 1 $\mu$as per star and assumed to be uncorrelated between field stars (and therefore averages down as $1/\sqrt{N_{star}}$ in the combined measurement).}
\end{figure}

Photon noise on field stars sets a fundamental limit to the astrometric accuracy, and selecting the few brightest stars in the field may therefore appear preferable, rather than optimizing the instrument for large number of fainter stars. Uncalibrated distortions and systematic errors will however require using many stars toward the measurement to average down non-photon noise errors. This is illustrated in Figure \ref{fig:noisefact}, which shows for a particularly simple example (fixed astrometric distortion residual equal to 1$\mu$as) how the two error terms (photon noise and non-photon noise) evolve as a function of the field stars stellar magnitude, assuming that only stars within 0.5 mag of a nominal stellar magnitude are selected. In an actual flight system, a wider range of brightness can be selected --- the narrower range is only adopted here to identify which stars contribute the most to the final accuracy. In this simple example, it is assumed that the 1 $\mu$as non-photon noise error per star is uncorrelated between stars. Selecting stars around $m_V=19$ then offers the best compromise between photon noise and distortions/systematic errors, yielding a single axis astrometric measurement accuracy below 0.1 $\mu$as in a 1 day observation for the 1 deg$^2$ (0.6 deg diameter) galactic pole field considered. With reduced systematic errors, brighter stars should be selected, even if they are less numerous, and the overall astrometric accuracy would be better. To estimate the final astrometric measurement error of our concept, it is therefore necessary to quantify measurement noise as a function of field star brightness and optimally select and combine field stars according to their apparent luminosities: this analysis is performed in this section using a numerical model which includes dominant sources of noise. 

When optical design and numerical simulations are required, we chose to adopt a medium-sized 1.4 m space telescope with a 0.3 deg$^2$ (0.6 deg diameter) FOV, although the technique could be applied to larger telescopes to offer higher astrometric accuracy. A numerical tool was developed to quantify the measurement accuracy and identify the main terms in the error budget. Section \ref{ssec:errterms} identifies possible sources of astrometric measurement error and, when possible, discusses their likely amplitude. These errors guide the numerical simulation, described in Section \ref{ssec:numsim}, which is designed to quantify the main error terms and produce an overall error budget. Table \ref{tab:errorterms} summarizes the error terms discussed and lists which ones are to be included in the numerical model.

\subsection{List of error terms considered}
\label{ssec:errterms}

\begin{deluxetable*}{llcl} 
\tabletypesize{\scriptsize}
\tablecolumns{4} 
\tablewidth{0pc} 
\tablecaption{\label{tab:errorterms}List of Error Terms Considered for the Numerical Model} 
\tablehead{ 
\colhead{} & \colhead{Description} & \colhead{Included in} & \colhead{Notes}\\
\colhead{} & \colhead{} & \colhead{Numerical Model ?} & \colhead{}}

\startdata 
\multicolumn{4}{c}{{\bf Photon noise and related effects} (see Section \ref{sssec:phnoise}) }\\
\hline
N1 & Photon noise (field stars)                  & Yes   & See the Appendix \\
N2 & Pixel sampling (field stars)                & Yes   & See the Appendix \\
N3 & PSF polychromaticity (field stars)          & Yes   & See the Appendix \\
N4 & Zodiacal light photon noise (field stars)   & Yes   & See the Appendix \\
N5 & Central star and zodi photon noise          & Yes   & Affects image of diffraction spikes\\
\hline
\multicolumn{4}{c}{{\bf Astronomical terms} (see Section \ref{sssec:astron}) }\\
\hline
N6 & Central star proper motion                  & Yes   & Fitted/removed in final measurements\\
N7 & Central star parallax motion                & Yes   & Fitted/removed in final measurements\\
N8 & Aberration of light                         & {\bf No}    & Effect is similar to, but much smaller than N6 + N7\\
N9 & Companions around field stars               & {\bf No}    & Averaged by large number of field stars\\
N10 & Central star photometric variability       & {\bf No}    & Small effect on spikes measurement\\
N11 & Stellar spots and activity                 & {\bf No}    & Expected to be smaller than 0.1 $\mu$as \\
\hline
\multicolumn{4}{c}{{\bf Detector} (see Section \ref{sssec:detec}) }\\
\hline
N12 & Uncalibrated errors in detector flat field          & Yes   & 1\% rms. 6\% peak\\
N13 & Uncalibrated detector spectral response errors      & Absorbed in N12   & Effect is absorbed in N12 numerical model \\
N14 & Intra-pixel detector sensitivity variations         & Absorbed in N12   & Effect is absorbed in N12 numerical model \\
N15 & Uncalibrated detector geometry error                & Yes   & Assumes uncalibrated temporal and spatial $\delta T \approx$ 20 mK \\

N16 & Variations in pixel sensitivities over time         & Yes   & Assumed to be at the 0.1\% level (excludes calibration)\\
N17 & Variations in detector geometry over time           & Yes   & Effect is included in N16 numerical model\\

N18 & Detector saturation                                 & Approximated    & Field stars brighter than $m_V = 14$ are excluded\\
N19 & Readout noise                                       & {\bf No}    & Exposure time chosen for photon-noise limited imaging\\
N20 & Uncalibrated detector non-linearity                 & {\bf No}    & No significant contribution expected in final measurement\\
N21 & Uncalibrated variations in readout timing      & {\bf No}    & $<$0.01 $\mu$as\\
\hline
\multicolumn{4}{c}{{\bf Telescope and optics} (see Section \ref{sssec:opttel}) }\\
\hline
N22 & Telescope pointing jitter                           & {\bf No}    & Negligible impact if below diffraction limit\\
N23 & Telescope roll angle errors                         & {\bf No}    & Fitted and removed from final data \\
N24 & Uncalibrated primary mirror surface errors          & {\bf No}    & Negligible impact if below diffraction limit\\
N25 & Uncalibrated static M2 and M3 figure errors & yes         & PSD of manufactured optics used for simulation\\
N26 & Telescope alignment errors                          & {\bf No}    & First order terms fitted, residual smaller than N6+N7\\
N27 & Plate scale error                                   & {\bf No}    & Less than 1e-3 $\mu$as\\
N28 & Local random errors in dot positions and size       & {\bf No}    & Non-significant if position error $< 10 \mu m$\\
N29 & Non-uniformity in dots coverage                     & {\bf No}    & Removed from measurement thanks to roll\\
N30 & Uncalibrated variation in M2 and M3 surface figure  & yes         & Produces astrometric distortion\\
\enddata 
\end{deluxetable*} 

\subsubsection{Photon noise and related effects}
\label{sssec:phnoise}

\paragraph{\bf N1 Photon noise on background stars}
This noise term limits the single axis achievable astrometric precision to $\sigma_1 = 1/(\pi \sqrt{N_{ph}})$ per field star, where $N_{ph}$ is the total number of photon acquired for the field star taking into account the telescope throughput and detector quantum efficiency.

\paragraph{\bf N2 Pixel sampling}
The equation given for noise N1 is only valid for a continuously sampled PSF. With flux integration on a grid of square pixels, the field star astrometric accuracy is degraded.

\paragraph{\bf N3 PSF polychromaticity}
The equation given for N1 also assumes a monochromatic PSF. A polychromatic PSF is smoother and, for the same number of photons, offers reduced astrometric accuracy.

\paragraph{\bf N4 Zodiacal light photon noise contribution}
The zodiacal background is a non-negligible source of photon noise for the fainter stars.

\paragraph{\bf N5 Photon noise in the diffraction spikes}
The finite number of photons in the diffraction spikes (which contain approximately 1\% of the total flux from the central star) and the zodiacal light background introduce photon noise which creates an error in the measurement of the distortion map used to compare observations at several epochs. This noise is uncorrelated with all other noise sources.

\paragraph{A quantitative description of terms N1, N2, N3 and N4 is given in the Appendix, and is included into the numerical model as an uncorrelated noise for each field star. N5 is included separately in the model when computing distortion maps from spike images.}

\subsubsection{Fundamental astronomical effects}
\label{sssec:astron}

\paragraph{\bf N6 Central star proper motion}
While the central star proper motion is not directly impacting the astrometric performance (it is a constant drift which is removed from the measurements), it will drag the field stars' positions on the sky relative to the telescope pointing. Static optical surface errors on M2 and M3 create astrometric distortions, and proper motion will move the field stars on different parts of this static distortion map for the different observation epochs. Even if the roll angle and optical surfaces are the same between observations, a time-variable astrometric error will be produced. The N6 term is therefore coupled to terms N15 and N25 described below. 

\paragraph{\bf N7 Central star parallax motion}
The effect of parallax motion is similar to proper motion: the field stars are moved on different parts of a static distortion map.

\paragraph{\bf N8 Aberration of light}
Aberration of light introduces a known time-variable stretch of the focal plane image, which is removed from the measurement. A second-order effect is that field stars are moved on slightly different parts of the field and will therefore pick up a slightly different astrometric distortion between epochs (this effect is similar, but much smaller in amplitude, to N6 and N7).

\paragraph{\bf N9 Companions around background stars}
Several of the field stars have companions, which will modulate their position. This error term is reduced thanks to the large number of stars used as an astrometric reference, and the fact that they are faint ($m_V > 14$) and relatively distant. We have not quantified this error term, and note that prior spectroscopic observations may be required to eliminate binary stars from the astrometric reference.

\paragraph{\bf N10 Central star photometric variability}
Photometric variability of the central star, coupled with non linear response of the detector and uncalibrated flat-field errors, will produce a small apparent change in the position of the diffraction spikes.

\paragraph{\bf N11 Stellar spots}
Stellar spots are a significant source of noise, and can produce astrometric errors of up to about 0.1 $\mu$as.

\subsubsection{Detector terms (static)}
\label{sssec:detec}

The focal plane detector array static errors create errors for both the measurement of spikes position and the background reference stars positions. While the detector errors considered here are static (they do not change between observing epochs), they still contribute to the final astrometric measurement errors due to the motion of background reference stars between observations (parallax. proper motion), and, to a lesser degree, to the small motion of the spikes due to variable optical distortions.

\paragraph{\bf N12 Uncalibrated errors in flat field response}
Static residual (after calibration) errors in the detector response from pixel to pixel are expected at the percent level. This will create a position-dependent error in the measured position of field stars.

\paragraph{\bf N13 Uncalibrated spectral response error}
Each pixel has a slightly different spectral sensitivity, due to manufacturing defects in the detector coating or variation in the substrate thickness (fringing in CCD cameras). 

\paragraph{\bf N14 Intra-pixel sensitivity variations}
While N12 and N13 describe the sensitivity averaged across each pixel, the sensitivity also varies within a single pixel, and this sub-pixel variation may not be identical between pixels.

\paragraph{\bf N15 Uncalibrated detector geometry error}
The pixel geometry of the detector is not perfectly known, producing a position-dependent astrometric error.

\subsubsection{Detector terms (time variable)}

Variation of the detector properties (geometry and response) between observations affects the position measurement of both the spikes and the background stars.

\paragraph{\bf N16 Variations in pixel sensitivities over time}
The pixel response can change with time, and this change can have a spectral and spatial (within each pixel) component. While an overall change in detector response has no effect on the astrometric measurement, differential variations between pixels will produce an error in the position measurement for both diffraction spikes and field stars. The contribution of this error term is small for field stars thanks to the averaging due to telescope roll, but it is more significant for the spikes which are fixed on the detector. 

\paragraph{\bf N17 Variations in detector geometry over time}
Temperature variations across the focal plane array will produce an astrometric distortion. If this distortion is different between observation epochs, it will contribute to the final astrometric measurement error.

\subsubsection{Other detector terms}

\paragraph{\bf N18 Detector saturation}
Detector saturation prevents bright field stars from being used for astrometric measurement. We note that while the bright field stars are theoretically the best choice in a photon noise limited measurement, they are far less numerous than faint stars, and therefore contribute little to the astrometric measurement in a distortion limited measurement.

\paragraph{\bf N19 Readout noise}
Detector readout noise introduces an error in the position measurement of both field stars and diffraction spikes.

\paragraph{\bf N20 Uncalibrated detector non-linearity}
Detector non-linearity does not contribute to the error budget if it is common to all pixels and constant in time. 

\paragraph{\bf N21 Uncalibrated variable errors in detector readout timing}
With the telescope continuously rolling, the position of field stars on the telescope is a function of time, while the diffraction spikes are static. An error in the readout timing therefore yields an error in the final astrometric measurement. With a 1 rad roll angle for a 1 day observation, a 100 $\mu$s error corresponds to a 1 $\mu$as error in the position of a field star at the edge of 0.5 deg diameter field. While a static readout timing error does not contribute to the astrometric error in our differential measurement, an unknown change in the detector timing readout between observations would produce an astrometric error. With several hundred stars participating to the final measurement, and many measurements at different roll angles, this error is averaged down. Assuming an error per field star corresponding to 10 $\mu$s = 0.1 $\mu$as at the edge of the field (which itself would be the roll average of a much larger error per individual exposure), and 100 stars, the final error would be 0.01 $\mu$as if all stars were at the edge of the field, and less for a realistic distribution of stars in the field.

\paragraph{Saturation (N18) is included in our model by removing the bright field stars from the astrometric measurement. Detector readout noise (N19) is not included in our model, as we assume that the readout time would be chosen for photon noise limited sensitivity. Detector non linearity (N20) is assumed to be smaller than N13 and N14.}

\subsubsection{Optics, telescope}
\label{sssec:opttel}

\paragraph{\bf N22 Telescope pointing}
Telescope pointing errors will affect equally field stars and diffraction spikes, and will therefore not produce first-order astrometric errors. Pointing errors can however impact astrometric accuracy by broadening the image of the field stars and spikes, yielding larger errors than predicted by the equation given for noise term N1. Pointing errors can also be aliased by the detector readout timing, and can produce an astrometric error if the image of the spikes and field stars are not simultaneous.

\paragraph{\bf N23 Telescope roll angle error}
The telescope roll angle will elongate PSFs proportionally to both exposure time and distance from the optical axis.

\paragraph{\bf N24 Uncalibrated M1 surface errors}
M1 surface errors will affect equally field stars and diffraction spikes, and will therefore not produce first-order astrometric errors. Large M1 errors can broaden the field star PSFs and spikes, impacting astrometric sensitivity. 

\paragraph{\bf N25 Uncalibrated static M2 and M3 surface errors}
Errors in the optical surface of M2 and M3 will produce a static unknown astrometric distortions due to beam walk on these optical elements. We note that surface errors on the PM do not produce such distortions, as it is the aperture stop of the system. Proper motion and parallax motion of the central star will slowly move the field stars on different parts of this distortion map at different observing epoch, therefore producing an astrometric error in the differential astrometric measurement between epochs. While the distortion calibration using the images of the diffraction spikes removes most of the low-order distortions, it does not efficiently correct for errors on a scale of the parallax + proper motion of the star, which is smaller than the separation between spikes. 

\paragraph{\bf N26 Telescope alignment errors}
Misalignments of the optics of the telescope can produce astrometric distortions. To first order (ignoring errors in the surface figure of the mirrors), these distortions are well known and can be removed from the measured set of 2D field stars positions at very little cost in sensitivity thanks to the large number of 2D field stars and the small number of distortion modes introduced by misalignments. These distortions are also very well calibrated by the diffraction spikes as they exhibit no high spatial frequency component in the focal plane. A second-order effect of misalignment is to introduce high spatial frequency distortions due to beam walk on aberrated M2 and M3 surfaces. We have computed this effect through high-resolution ray tracing through the optical system, repeated with different rigid body positions and angles of M2 and M3. This effect is smaller than the beam walk effect due to proper motion and parallax motion for realistic optics displacements (few tens of $\mu$m) and tilts (few arcseconds): on M3, the optical element most sensitive to beam walk, the $\approx$ 1 arcsec motion on the sky due to proper motion and parallax corresponds to more than 0.1 mm or 2.5 arcsec of mechanical tilt of M2. In a stable temperature-controlled telescope system using low coefficient of thermal expansion materials, rigid body motion of the optics can be kept smaller than these values.

\paragraph{\bf N27 Plate scale error}
To first order, an error in the plate scale does not affect the astrometric measurement, as the measurement is only performed in the angular coordinate (radial component is discarded). Plate scale variations are however multiplicatively coupled with proper motion and parallax motion, which are much larger than the signal to be measured. For example, with a 1 arcsec year$^{-1}$ proper motion, and a 1e-7 relative variation in the plate scale between epochs, the measured astrometric motion of the central star over one year contains a $10^{-7} \times$ 1 arcsec $=$ 0.1 $\mu$as error along the proper motion direction. The diffraction spikes do not allow an accurate measurement of the overall plate scale at the detector. Variations in the telescope alignment or overall thermal expansion of optics or the focal plane array can create plate scale variations between observation epochs. To remove this error term, field stars are used to independently measure plate scale at each epoch. A 1e-7 relative change in plate scale corresponds to 0.2 mas or 1/200 of a pixel over the 0.6 deg diameter field considered in our baseline design. Since the plate scale measurement relies on the same field stars as the astrometric measurement, and since the astrometric measurement (using the azimuthal coordinate) achieves 0.2$\mu$as precision, a 1000 times larger 0.2 mas precision should be easily achievable using the radial coordinate. We however note that a key difference between the plate scale measurement and the astrometric measurement is that the beneficial roll anticorrelation effect described in Section \ref{sec:telescoperoll} only applies to the azimuthal measurement. Accurate plate scale measurement may therefore rely on a larger number of fainter stars to average down detector errors.

\paragraph{\bf N28 Local random errors in dot positions and size}
Uncorrelated errors in the position, size or shape of the dots on the PM will broaden the spike images, leading to reduced sensitivity of the distortion measurement.

\paragraph{\bf N29 Non-uniformity in dots coverage}
Errors in the position or size of dots which are correlated over large distances will create an effective non-uniformity on the dot coverage $\tau$ (locally equal to the geometrical fraction of the mirror covered by the dots). $\tau$ then becomes a function of position on the mirror, and differs from $\tau_{ave}$ the average value of $\tau$ over the mirror. While the position of field stars is a linear function of the wavefront slope across the PM weighted by the pupil illumination $\alpha_{fs}$ after light has been removed by the dots ($\alpha_{fs} = (1 - 2 \tau)/(1 - 2 \tau_{ave})$), the position of the spikes is a linear function of the wavefront slope weighted by the dot coverage $\alpha_{sp} = \tau/\tau_{ave}$. For example, a focus error between observations corresponding to a 10 nm rms wave front error, coupled with a 0.1\% gradient in the relative dot coverage from one edge of the pupil to the opposite edge, will create a 2.4 $\mu$as shift between the diffraction spikes and the field stars (this number is computed by multiplying the wavefront focus phase function with the non-uniform amplitude containing the gradient and measuring the tip-tilt component of this product averaged over the pupil). Since our final astrometric measurement is differential, a fixed wave front error on the PM coupled with a fixed error in the dot coverage will not lead to an astrometric measurement error. However, a change in the figure of the PM between observation epochs will lead to an astrometric error. Assuming a static wavefront error during each epoch, the error is a fixed vector in the telescope/instrument coordinate system, while the astrometric signal to be measured is a fixed signal in sky coordinates. If the telescope were to be rolled by 360 deg during a single observation epoch, this error would therefore disappear, as the astrometric shift between the spikes and the field star will rotate on the sky with telescope roll angle, and will average to zero for a full telescope rotation. For a 180 deg rotation, the 2D shift would be averaged to zero in one direction, and attenuated by $\int_0^{\pi} \sin(x) dx/\pi = 1/\pi$ in the other direction. This error can also be filtered out of the astrometric measurement by rejecting the part of the measurement which is correlated with roll angle in the way described above (and scale up the remaining part of the measurement to account for the fact that this operation will also attenuate the final measured astrometric value). While in theory this would eliminate the problem for any non-null roll angle, its cost in required signal-to-noise ratio (SNR) is prohibitive when the measurement is done over a small range of roll angles, and can only be effective for roll angles of about 90 deg or more. If no such fitting is done, and with a 1 nm rms change in the average mirror figure between observations, the above analysis suggests that the dots need to be uniform to 0.04\% from one side of the mirror to the opposite side for this error term to be below 0.1 $\mu$as. If this non-uniformity is entirely allocated to dot size, the gradient in the diameter of the dots should be less than 2\% from edge to edge of the PM. Higher order and more localized spatial variations of the dot coverage can be larger, as they will be less correlated with expected changes in the mirror figure, which are larger at low spatial frequencies.

\paragraph{\bf N30 Uncalibrated variations in M2 and M3 surface figure}
As the optical surface of M2 and M3 changes between observation epochs, so does the astrometric distortion in the focal plane. While the diffraction spikes allow for calibration of most of this effect, this calibration is not exact due to the finite sensitivity and spatial sampling of the distortion measurement with the spikes.

\subsection{Numerical simulation approach}
\label{ssec:numsim}

\begin{deluxetable*}{lccl} 
\tabletypesize{\scriptsize}
\tablecolumns{4} 
\tablewidth{0pc} 
\tablecaption{\label{tab:baselinedesign} System Parameters for the Baseline Design and the Numerical Simulation} 
\tablehead{ 
\colhead{} & \colhead{Baseline} & \colhead{Numerically} & \colhead{Rationale} \\
\colhead{} & \colhead{Design}   & \colhead{Simulated}   & \colhead{} }

\startdata 
\hline
Telescope diameter & \multicolumn{2}{c}{1.4 m} & Assumed (cost constrained)    \\
Detector pixel size & \multicolumn{2}{c}{44 mas} & Nyquist sampled at 600 nm     \\
Field of view (FOV) & 0.29 deg$^2$ & 0.03 deg$^2$ & Set to meet astrometric accuracy requirement    \\
                    & (0.6 deg diam) & (0.2 deg diam) & Optical design allows diffraction-limited imaging over FOV   \\
Single measurement duration & \multicolumn{2}{c}{48 hr} & Typical single coronagraphic observation duration \\
Dot coverage on PM (area) & 1\% & 0.12\% & Keeps throughput loss moderate \\
Dot size / pitch ($\mu$m) & 120/932  & 360/2800  & dot diameter imposed by FOV \\
                          & (black dots) & (gray dots) & \\ 
Flat-field error, static & \multicolumn{2}{c}{1.02\% rms, 6\% peak} & Conservative estimate for modern detectors \\
Flat-field error, dynamic & \multicolumn{2}{c}{0.1\% rms pixel$^{-1}$} & Undetected cosmic-ray damage on detector \\
                          & \multicolumn{2}{c}{Spatially uncorrelated} & \\
Detector distortion, static & \multicolumn{2}{c}{0.2\% of pixel size} & \\
Telescope roll (full range) & $\pm$ 10 deg & $\pm$ 28 deg & Manageable PSF elongation at edge of FOV\\
M2/M3 surface change & \multicolumn{2}{c}{40 pm} & Level of surface deformation experienced in optical labs\\
M2/M3 surface error & \multicolumn{2}{c}{1.5 nm} & Surface error and PSD from existing optics\\
\enddata 
\end{deluxetable*}

\subsubsection{Overview}

\label{ssec:overview}
\begin{figure*}
\includegraphics[scale=0.90]{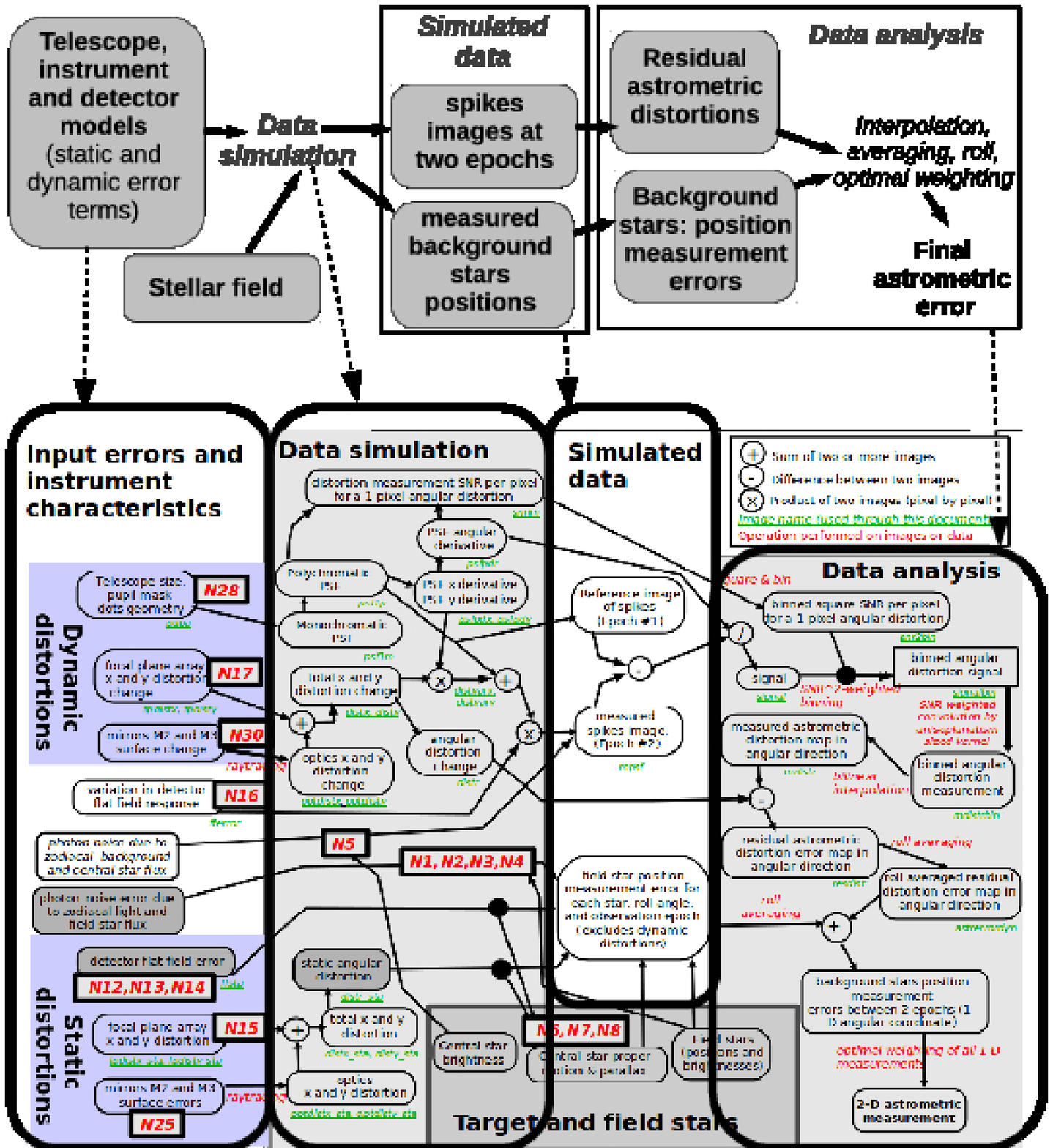} 
\caption{\label{fig:numsimulchart} 
Numerical simulation overview: overview (top) and detailed step-by-step description (bottom). The data acquired by the wide-field camera is first simulated, including the relevant sources of error listed in Table \ref{tab:errorterms} and discussed in Section \ref{ssec:errterms}. Images of the diffraction spikes and position measurement for the background stars are computed separately as discussed in the text. The right part of the figure shows how the simulated data (or actual data from a mission) are analyzed to produce the 2D astrometric measurement.}
\end{figure*}

A numerical simulation was developed to quantify the astrometric accuracy of the system. The top part of Figure \ref{fig:numsimulchart} gives an overview of the numerical simulation, and a more detailed step-by-step description is given in the bottom part of the same figure. The relevant sources of noise identified in Section \ref{ssec:errterms} are included in the model, and the detailed block diagram of the numerical simulation shows where and how they are inserted. 

The {\bf model input (telescope, instrument, and detector models)} is first established. As detailed in the bottom left block of Figure \ref{fig:numsimulchart}, this includes a model of the focal plane detector array distortion (static and dynamic), a model of the optical surface errors on mirrors M2 and M3 (static and dynamic), and variations in detector flat field. The main model parameters are also chosen such as telescope diameter and pupil mask geometry. The {\bf stellar field} is computed in coordinates relative to the central star, and is therefore translated between observation epochs according to proper motion and parallax (the actual astrometric signature of a planet is much smaller than proper motion and parallax, and therefore has no impact on the noise and astrometric measurement accuracy).

The {\bf data simulation} is performed using the model inputs and a model of the stellar field to be observed. As shown in the "data simulation" block in the bottom of Figure \ref{fig:numsimulchart}, this step relies on ray tracing to transform the optical surface errors into astrometric distortions, and applies these distortions (along with other defects such as detector flat field variations) to a simulated image of the diffraction spikes. Several sources of errors are also included to compute the set of measured background star positions, consisting of a 2D centroid in detector pixel units for each star, at each roll angle.

The {\bf simulated data} consist of two simulated diffraction spikes images (one with no source of noise or distortion, and one with distortions and noise included) and noisy sets of measured background star positions (one set for each observation epoch).

The {\bf data analysis} (bottom right box in the bottom of Figure \ref{fig:numsimulchart}) is performed by differentiating the two diffraction spikes images to first estimate the noise residual on the astrometric distortion measurement. This step includes a 2D interpolation of a signal which is only present along the 1D spikes to create a full continuous 2D map of angular distortion variations between two observation epochs. The residual distortion error is applied to the noisy pixel coordinate centroiding measurements previously computed. These sets of pixel coordinates are roll averaged and combined into a 2D astrometric measurement. This whole process is repeated for each observation epoch with different noise realizations.

The main parameters of the baseline system are listed in Table \ref{tab:baselinedesign}. The simulation is performed over a 0.1 deg radius field to keep the image size sufficiently small (16k pixels on a side, 1GB size per image in single precision) for fast computations. The results obtained with the small 0.03 deg$^2$ ($=$ 0.1 deg radius) field are then scaled to larger fields assuming that the final measurement error scales as the inverse square root of the FOV, provided that the surface brightness of the diffraction spikes is kept constant (this last requirement implies that scaling small FOV results to larger FOV also requires the fraction of the PM area covered by dots to scale linearly as the FOV). 

The model is not a true end-to-end simulation, as it was designed to be as simple as possible while accurately quantifying the relevant sources of errors identified in Table \ref{tab:errorterms} and discussed in Section \ref{ssec:errterms}. The numerical model does not produce complete simulated images as would be seen by the detector: diffraction spikes images are computed without field stars (top left part of the lower chart in Figure \ref{fig:numsimulchart}), and the distortions are measured from these diffraction spike images alone. Errors on the measurement of the position of field stars are treated separately, and added to the errors due to the imperfect calibration of distortions by the spikes. This simplification to the numerical simulation is made possible by the non-overlap between sources of errors affecting distortion calibration with the spikes (error terms N5, N16, N17, N25, and N28) and the errors affecting the measurement of the field stars locations in pixel coordinates (error terms N1, N2, N3, N4, N6, N7, N8, N12, N13, N14, N15, and N25).

\begin{figure*}
\includegraphics[scale=0.35]{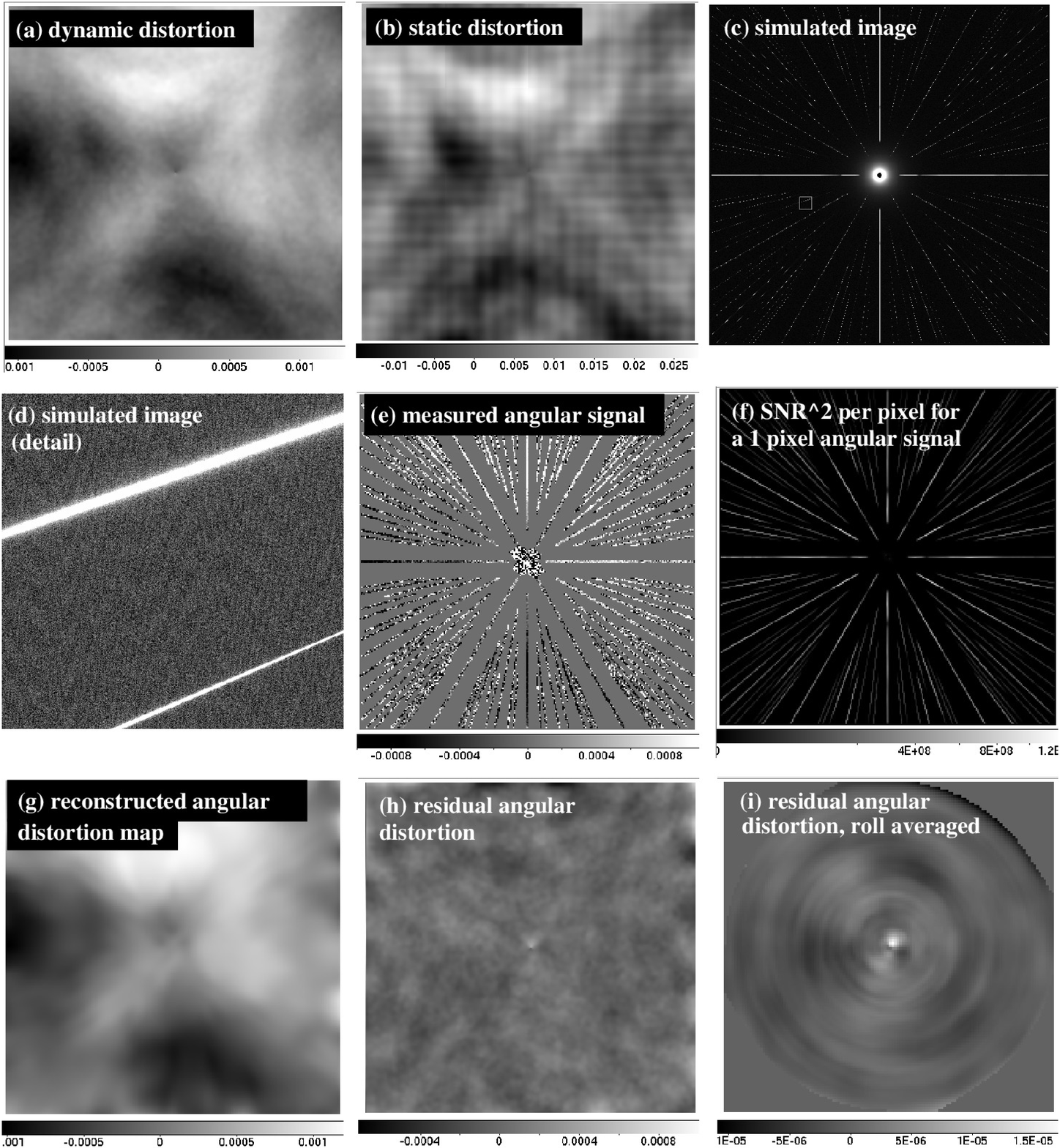} 
\caption{\label{fig:numsimexample} 
Frames extracted from key steps of the numerical simulation. Each image covers 0.2 deg $\times$ 0.2 deg on the sky, with a 44 mas pixel sampling. (a) Simulated distortion change between two observations in the angular direction, perpendicular to the spikes (unit $=$ pixel, image labeled $distr$ in figure \ref{fig:numsimulchart}). (b) Static distortion equal to the sum of static optical distortions and static detector distortion (unit $=$ pixel). This distortion, labeled $distr\_sta$ in Figure \ref{fig:numsimulchart}, is common to all observations, and is not removed by the calibration using diffraction spikes. The static distortion (b) includes detector manufacturing effects at high spatial frequency, and is therefore less smooth than the dynamic distortion. (c) Simulated image, without background field stars, showing the diffraction spikes. (d) Simulated image detail: the area shown in this image is the small white square in (c) to the bottom left of the center. Photon noise from zodiacal light background is visible. (e): The measured angular signal is the estimated displacement in pixel, of the diffraction spikes in the angular direction (perpendicular to the spikes). It is computed for each pixel and is shown here for pixels whith sufficient illumination from the diffraction spikes. (f) Squared signal-to-noise for a 1 pixel angular displacement of the spikes, shown for each pixel of the image. (g) By interpolating (e) using weighting given by (f), a full map of the angular distortion change is reconstructed. (h) The residual angular distortion after calibration is equal to (a)--(g), and shown here. (i) The roll-averaged angular distortion shows for each position on the sky the residual angular distortion error after roll averaging (unit $=$ arcsecond); it is at the few $\mu$as level, and, when averaged between several field stars, yields the final sub-$\mu$as astrometric measurement accuracy.}
\end{figure*}

Key steps of the numerical simulation are shown in Figure \ref{fig:numsimexample}. The figure shows static and dynamic distortion maps computed with realistic assumptions about mirror shapes, deformations and focal plane array shape and stability. A power spectral density measured on previously manufactured mirrors of similar size, curvature, and asphericity was kindly provided by L-3 Tinsley and used for this model. The mirror deformations between observations are assumed to be 40 pm per surface, which is similar to what is experienced on mirrors of this size in optical manufacturing laboratories with $~$0.25$^{\circ}$C temperature control (with mK-level temperature control on orbit, the deformations would likely be much smaller). It is assumed that the true pixel location differs from expectation by 0.2\%, with most of this difference at spatial frequencies higher than can be sampled by the diffraction spikes. We note that this is a conservative assumption, and that such geometry errors could be calibrated prior to launch. The focal plane array is assumed to have uncalibrated temperature inhomogeneities varying by 20 mK between observations, and the coefficient of thermal expansion of silicon is assumed to convert these temperature differences into deformations of the focal plane. With these assumptions, a raytrace model of the optical design is used to simulate both the static distortion and its change (dynamic) between observations, as shown by images (a) and (b) in Figure \ref{fig:numsimexample}. A pair of simulated images of the central bright star as seen by the wide-field camera can then be computed: one with the static distortion only, and one with the static and dynamic distortion terms (images c and d). Differentiating the two images produces the measured angular distortion signal (e), which only contains a high SNR measurement along the diffraction spikes (f). This signal is then interpolated into a 2D map of measured distortion change between the two observations (g), which is subtracted from the true distortion change to produce the residual error in distortion (h). The telescope roll averages this residual, and the resulting map (i) shows, for each possible field star position in the field, the error due to uncalibrated changes in distortion.

Our numerical model makes a number of conservative assumptions, often motivated by the desire to keep the numerical simulation simple.
\begin{itemize}
\item{Galactic pole pointing is assumed, resulting in a low density of background field stars. Only stars are used for astrometric referencing: extragalactic objects, which outnumber stars at $m_V=20$ and fainter, are not considered, and would help with astrometric referencing at high galactic latitude.}
\item{We assume no calibration of static errors of the system prior to launch. The amplitude of optical and detector errors adopted in our model corresponds to currently available hardware, but we assume no knowledge of these errors. This is a highly conservative assumption, as ground testing can measure these errors which can then be taken into account in the data analysis. For example, optical surfaces of M2 and M3 would likely be available from the manufacturer, providing a good estimate of static optical distortions. Detector errors (flat field and regularity of pixel positions) can also be measured prior to launch.}
\item{We assume no on-orbit calibration of error, and no attempt to correlate errors with system variables (such as spacecraft Sun angle, target star stellar type) is made. We assume that each observation is analyzed with no knowledge of other observations. This is a very conservative assumption, as static distortions can be partially calibrated by dithered observations (see, for example, \cite{2011PASP..123..622B}) and changes in the optical system such as optics deformation are likely to be either gradual (can be measured using the previous and next observations), or correlated to identifiable system variables (for example, spacecraft Sun angle), and can be at least partially compensated for by analyzing globally all measurements, including measurements on other targets. We also note that dominant error sources (for example, due to temperature changes in the telescope structure or intermittent detector effects), even if they cannot be predicted, modeled, or linked to a physical process, will produce recognizable signatures in the data, which can be identified by processing all data acquired during the mission and removed from the data prior to the final astrometric measurement. This is made possible by the large redundancy in the measurement, where the positions of many stars, over many locations of the focal plane, are used to compute a few variables (planet(s) orbital parameters and mass(es)). Past and future astrometric missions rely on such techniques to reduce astrometric errors, but for simplicity, we have chosen to not do this, as it is difficult to quantify how much this global analysis of the measurements improves performance without detailed simulations of the disturbances in the telescope and instrument.}
\end{itemize}

Our simple model also makes a number of optimistic assumptions.
\begin{itemize}
\item{A 100\% duty cycle is assumed for the measurement, and the detector readout is assumed to be instantaneous.}
\item{Several detector effects and errors are not included such as cosmic-ray hits and detector blooming of saturated stars in the field. Detector readout noise is not included in the simulation, and it is therefore assumed that the exposure time is chosen to operate in the photon noise limited regime for the field star brightness range offering the best astrometric signal. Charge diffusion and interpixel capacitance are not taken into account, and can contribute to the detector spatial modulation transfer function, which we assume here is dominated by the pixel size and geometry.}
\item{As described earlier in this section, we compute images of the diffraction spikes without field stars for calibration of astrometric distortions. This assumes that the data analysis can perfectly separate diffraction spikes from field stars. While this assumption is optimistic, we note that discarding parts of each frame where field stars and spikes overlap would result in a very small loss of sensitivity, as field stars are sparse on the focal plane array.}
\item{We have assumed that the diffraction spikes are fixed on the focal plane array between observing epochs, and that the distortion change between epochs is measured by comparing the intensity over the same pixels. This approximation implies that static uncalibrated errors in the detector flat field (term N12) have no effect on the distortion measurement, and that variations in pixel sensitivities over time (term N16) are the dominant non-photon noise error term. Figure \ref{fig:numsimexample} (image (a)) does indicate that variations in optical surfaces will produce $10^{-3}$-level variations in the location of the spikes, and the focal plane detector array distortions (excluding translation and focus) could be stable to a few $10^{-3}$ pixel between epochs (a 100 mK change in temperature non-uniformity on a detector array with 25,000 pixel across the field radius will produce a shift under $10^{-3}$ pixel at the edge of the field). The assumption that spikes are static on the detector therefore seems reasonable at the few $10^{-3}$ pixel level, assuming that the telescope pointing is driven to position the diffraction spikes on a fixed location, which removes the first-order effect of optical alignment errors. We note that even if the spikes were moved between epochs by a pixel or more, pixelation errors at the 1/100 pixel level per $\lambda/D$-long spike segment (corresponding to percent-level uncalibrated pixel responses) would be averaged down to approximately $0.1 \mu$as thanks the number of spikes ($\approx 100$) and their length ($\approx$ 10,000 resolution elements).}
\end{itemize}

Misalignments between the optical elements of the telescopes are not considered in the error budget, but are discussed separately in Section \ref{ssec:telalign}, where tolerances on the rigid body motions of optics are estimated.

Finally, we do not consider in our model variations of the astrometric distortions during a single observation. Our error budget only considers variation between different observations of the same source. This simplification should not affect the final astrometric measurement, as variations during a single measurement are averaged through the telescope roll, even if this averaging is, in the case of field stars, both spatial and temporal due to the roll of the telescope.


\subsection{Results and Error Budget}

\begin{deluxetable}{lcc}
\tablecolumns{3} 
\tablewidth{0pc} 
\tablecaption{\label{tab:errorbudg} Error Budget for 0.29 deg$^2$ Field, Galactic Pole} 
\tablehead{ 
\colhead{Error Term} & \colhead{Value} & \colhead{Notes}} 

\startdata 
\multicolumn{3}{c}{\bf Photon noise}\\
Field stars' centroid error & 0.128 $\mu$as & N1+N2+N3+N4\\
Photon noise on spikes & 0.048 $\mu$as & N5\\
\hline
\multicolumn{3}{c}{\bf Static calibration errors}\\
Detector flat field & 0.033 $\mu$as & N12+N13+N14\\
Optical distortion & 0.083 $\mu$as & N6+N7+N8, N25\\
Detector distortion & 0.015 $\mu$as & N15, N25\\
\hline
\multicolumn{3}{c}{\bf Dynamic calibration errors}\\
Detector flat field  & 0.029 $\mu$as & N16\\
Optical distortion  & 0.063 $\mu$as & N30\\
Detector geometry  & 0.076 $\mu$as & N17\\
\hline
{\bf Total} & 0.200 $\mu$as& \\
\enddata 
\end{deluxetable} 

\begin{figure}
\includegraphics[scale=0.45]{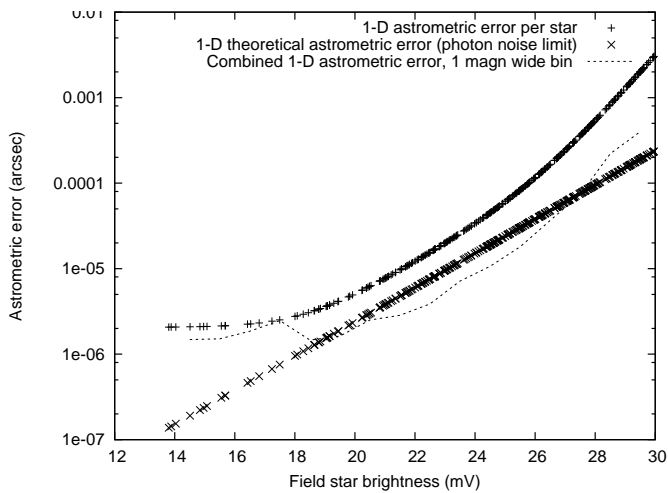} 
\caption{\label{fig:astrerr_per_star} 
Astrometric 1D error per star on the narrow 0.03 deg$^2$ ($=$ 0.2 deg diameter) FOV numerically simulated. Each point corresponds to a star in the field. Stars brighter than $m_V=14$ have been excluded, as they are assumed to be saturated on the detector. The straight line of points shows the expected astrometric error from photon noise alone in monochromatic light ($\lambda=0.5 \mu$m) with infinite pixel sampling and no background. The curved line of points shows the achieved measurement precision, and differs significantly from the photon noise limit both at the bright and faint ends. The dashed curve shows the final 1D astrometric accuracy obtained if only stars in a 1 mag wide brightness bin are used.}
\end{figure}

The numerical simulation results, summarized in the error budget in Table \ref{tab:errorbudg}, show that a 0.2 $\mu$as astrometric accuracy (1 $\sigma$ error per measurement per axis for a 2 day observation) is achieved with a 0.29 deg$^2$ ($=$ 0.6 deg diameter) FOV. The four dominant sources of error are photon noise on the field stars (0.128 $\mu$as), static optical distortions (0.083 $\mu$as), changes in detector geometry (0.076 $\mu$as), and changes in optical distortions (0.063 $\mu$as). Together, these four terms contribute to more than 90\% of the total astrometric error. Photon noise on the diffraction spikes (0.048 $\mu$as), detector flat field errors (0.033 $\mu$as), static detector distortion (0.015 $\mu$as), and changes in the detector flat field (0.029 $\mu$as) are all much smaller.

The fact that the overall contribution of photon noise on field stars (0.128 $\mu$as) is comparable to the sum of all other error terms (0.154 $\mu$as) is due to the weighting of signals from individual stars in the algorithm. With large calibration errors, the weighting favors a large number of faint stars rather than a small number of bright stars, which would yield a small photon noise error but a large calibration error due to poor $\sqrt{N}$ averaging. The weighting therefore automatically balances the two error terms, always yielding a field star photon noise contribution approximately equal to the calibration error in the error budget. This effect is described in Section \ref{sec:intro} and illustrated by Figure \ref{fig:noisefact}. For the small 0.03 deg$^2$ ($=$ 0.2 deg diameter) field simulated, Figure \ref{fig:astrerr_per_star} shows the 1D astrometric error per field star. The final astrometric measurement is obtained by combining 1D measurements obtained on the field stars with weighting factors chosen to minimize the final error ($SNR^2$ weighting). Figure \ref{fig:astrerr_per_star} shows that the astrometric error is smallest for bright stars, but never goes below $\approx$ 2 $\mu$as due to calibration residuals. For stars fainter than $m_V \approx 18$, photon noise related errors dominate, and the error therefore grows rapidly with stellar magnitude. The figure also shows that, if only stars within a 1 mag wide bins were considered, the best astrometric measurement is obtained by selecting stars around $m_V \approx 19$, as this brightness offers the best compromise between number of stars and measurement error per star. Between $m_V \approx 14$ and $m_V \approx 22$, each magnitude bin contributes almost equally to the final astrometric precision, as the decrease in precision per star as stellar magnitude increases is compensated by the increase in numbers of stars. 

Figure \ref{fig:astrerr_per_star} also shows the theoretical monochromatic astrometric measurement error per star, with the corresponding points aligned on a straight line in the log--log scale. This level of error is achieved if detector spatial sampling is infinite, in monochromatic light in the absence of calibration errors or background. At the bright end of the figure, the achieved error is much larger than this limit due to calibration errors. At the faint end, photon noise of the zodiacal background light dominates the error, as shown by the strong deviation of the curve from the ideal straight line. Even at the intermediate brightness where both zodiacal light photon noise and calibration errors are small compared to the theoretical limit, the achieved measurement error is about twice as large as the $1/(\pi \sqrt{N_{ph}} \lambda/D$ photon noise ideal limit, due to limited pixel sampling and polychromaticity of the PSF.

Figure \ref{fig:astrerr_per_star} illustrates the robustness of the concept performance against moderate increases of the sources of calibration noise. If the calibration floor were to be multiplied by 2 (4 $\mu$as instead of 2 $\mu$as), the final astrometric measurement error would only increase by a 23\% to 0.246 $\mu$as, as a larger number of fainter stars would be selected to mitigate the increased calibration error.
The final performance is therefore not highly sensitive to changes in the assumptions made in this paper about optical quality, detector properties and instrument stability. This suggests that obtaining substantially better accuracy than computed in this section would require improvements in the quality and stability of the system which are significantly beyond the realistic assumptions made in this paper. Better astrometric accuracy is therefore likely to require a combination of larger telescope or FOV (discussed in Section \ref{ssec:perfDFOV}) or modifications of the concept aimed at calibrating some of the dominant sources of calibration error.

\subsection{Performance as a Function of Aperture Size and Field of View}
\label{ssec:perfDFOV}

\begin{deluxetable}{lcccc}
\tablecolumns{5} 
\tablewidth{0pc} 
\tablecaption{\label{tab:perfdiamfov} Expected Single Measurement Astrometric Accuracy as a Function of Telescope Diameter and Field of View (Galactic Pole Pointing, 2 Day Integration)} 
\tablehead{ 
\colhead{\bf Tel. } & \multicolumn{4}{c}{{\bf Field of View}}\\
\colhead{\bf Diam.} & \colhead{0.03 deg$^2$} & \colhead{0.1 deg$^2$} & \colhead{0.3 deg$^2$} & \colhead{1 deg$^2$}\\
\colhead{} & \colhead{(\o$=$0.2 deg)} & \colhead{(\o$=$0.36 deg)} & \colhead{(\o$=$0.6 deg)} & \colhead{(\o$=$1.13 deg)}
}

\startdata 
1.0 m & 0.99 $\mu$as & 0.54 $\mu$as & 0.31 $\mu$as & 0.17 $\mu$as\\
1.4 m & 0.62 $\mu$as & 0.34 $\mu$as & 0.20 $\mu$as & 0.11 $\mu$as\\
2.0 m & 0.38 $\mu$as & 0.21 $\mu$as & 0.12 $\mu$as & 0.066 $\mu$as\\
2.8 m & 0.24 $\mu$as & 0.13 $\mu$as & 0.076 $\mu$as & 0.041 $\mu$as\\
4.0 m & 0.15 $\mu$as & 0.081 $\mu$as & 0.047 $\mu$as & 0.026 $\mu$as\\
5.7 m & 0.092 $\mu$as & 0.050 $\mu$as & 0.029 $\mu$as & 0.016 $\mu$as\\
8.0 m & 0.059 $\mu$as & 0.032 $\mu$as & 0.019 $\mu$as & 0.010 $\mu$as\\
\enddata 
\end{deluxetable}

The astrometric measurement error is estimated for different values of telescope diameter and FOV in Table \ref{tab:perfdiamfov}. The numbers shown in the table were derived by separating, for each star in the field, the error into a photon noise term and a calibration term. The photon noise term includes photon noise from the field star and zodiacal light, and the calibration term is a fixed error per star. The quadratic sum of the two terms matches for the nominal design ($D$ = 1.4 m, FOV = 0.3 deg$^2$ = 0.6 deg diameter), the curve obtained in Figure \ref{fig:astrerr_per_star}, and therefore takes into account all sources of error described in this paper. To evaluate how measurement error varies with telescope diameter, we scale for each background star the photon noise error with telescope diameter (this scaling is done separately for photon noise due to zodiacal light and field star photon noise), and we assume that the calibration residual error per star scales as the inverse of telescope diameter. This last assumption is somewhat arbitrary, but can be justified by the fact that the pixel size scales as $\lambda/D$, and that the width of the diffraction spikes used for calibration also scales as $\lambda/D$. We however note that proper extrapolation of calibration residuals for different telescope diameters requires optomechanical designs for different telescope sizes, and is beyond the scope of this introductory paper. We also note that the results shown in Table \ref{tab:perfdiamfov} are not strongly dependent on the calibration error per star, but are in large part driven by the ability to use a large number of fainter stars with the larger telescope sizes, and that this latter effect is well understood and quantified by simple scaling laws.

While the measurement is limited by systematics (scaling as $\lambda/D$) for bright field stars, it is limited by photon noise (scaling as $\lambda/D^2$) for fainter stars. For a fixed FOV, the overall measurement error dependence on telescope diameter is therefore intermediate between these two scaling laws: with larger telescope diameters, a larger number of fainter stars is used toward the final measurement to overcome the systematics and take advantage of the lower photon noise.

\subsection{Alignment Stability of Telescope Optics}
\label{ssec:telalign}

\begin{deluxetable*}{lccl}[t] 
\tablecolumns{3} 
\tablewidth{0pc} 
\tablecaption{\label{tab:telalign} Telescope Optics Alignment Sensitivity and Tolerances for Astrometric Imaging} 
\tablehead{ 
\colhead{Alignment Error} & \colhead{Alignment Sensitivity}         & \colhead{Alignment Tolerance for 0.1 pixel} & \colhead{Notes} \\
\colhead{}                & \colhead{(rms Angular Distortion)} & \colhead{rms Angular Distortion} & \colhead{}}
\startdata 

M2, $x$ translation          & 502 pixel m$^{-1}$      & 199 $\mu$m & $X$-axis is vertical in Figure \ref{fig:optprinciple}\\ 
M2, $y$ translation          & 961 pixel m$^{-1}$      & 104 $\mu$m & $Y$-axis is perp. to the plane of Figure \ref{fig:optprinciple}\\ 
M2, $z$ translation          & 443 pixel m$^{-1}$      & 226 $\mu$m & $Z$-axis is horizontal in Figure \ref{fig:optprinciple}\\ 
M2, rotation around $x$-axis & 3.21e3 pixel rad$^{-1}$ & 6.42 arcsec & Rotation axis intersects conic vertex\\ 
M2, rotation around $y$-axis & 3.15e3 pixel rad$^{-1}$ & 6.55 arcsec & Rotation axis intersects conic vertex\\ 

M3, $x$ translation          & 763 pixel m$^{-1}$      & 131 $\mu$m & \\ 
M3, $y$ translation          & 639 pixel m$^{-1}$      & 156 $\mu$m & \\ 
M3, $z$ translation          & 89.7 pixel m$^{-1}$     & 1.11 mm  & \\ 
M3, rotation around $x$-axis & 997  pixel rad$^{-1}$    & 20.7 arcsec  & Rotation axis intersects conic vertex\\ 
M3, rotation around $y$-axis & 1.96e3 pixel rad$^{-1}$ & 10.5 arcsec  & Rotation axis intersects conic vertex\\ 

Detector, z translation    & 62.1 pixel m$^{-1}$      & 1.61 mm & \\ 

\enddata 
\end{deluxetable*} 

\begin{figure}
\includegraphics[scale=0.46]{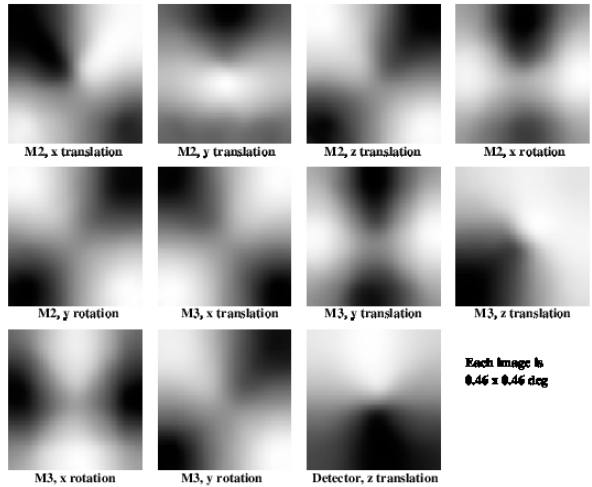} 
\caption{\label{fig:telalign} 
Amplitude maps for the angular component of astrometric distortions due to misalignment of optics. Each map shows a 0.46 $\times$ 0.46 deg field, and the linearly grayscale value represents the displacement of diffraction spikes or field stars in the angular (perpendicular to the spikes) direction. Each brightness scale is different, as shown by the different value for the rms angular distortions in Table \ref{tab:telalign}}
\end{figure}

The analysis presented in this paper has assumed so far a perfectly stable telescope alignment, and we address in this section the validity of this assumption.

\subsubsection{Astrometric Distortion Generated by Optical Misalignments}
Rigid body motions of the telescope optics create variations in the instrumental distortion, affecting the position of both the diffraction spikes and the field stars on the focal plane detector array.

The effect of rigid body motions of optical elements on diffraction spikes motion is quantified in Table \ref{tab:telalign}, and the morphology of these distortions is shown in Figure \ref{fig:telalign}. For each alignment error (rotation or translation of an individual component), high accuracy raytracing was used to compute astrometric distortion variation, with $\approx$ 3e6 rays per source and 1000 sources across the FOV. For each source, ray tracing was performed with and without the alignment error, and the difference between the locations of the two images in the focal plane constitutes the astrometric distortion variation. The average value of this distortion across the field (global pointing offset) is first removed, as the telescope would be pointed to locate the spikes on the same pixels between observations. The global roll term (rotating the spikes pattern) is not removed, as it is assumed that the focal plane array cannot be rotated against the telescope PM. The residual distortion is then measured in the angular coordinate only (the spikes are radially elongated, and a radial shift of their positions does not have an effect on the measurement). The values given in the second column of Table \ref{tab:telalign} are, for each mode, the angular distortion variation (rms averaged across the field) in detector pixel per unit of rigid body motion. For example, a 1 mm translation of M2 along the $x$ axis will move the diffraction spikes by 0.502 pixel rms in the focal plane. We have assumed for this table the baseline design (1.4 m telescope, 44 mas pixel size) used in this paper. The third column gives for each alignment mode the amount of rigid body motion required to produce a 0.1 pixel rms motion of the spikes, assuming that only a single misalignment mode is present.

\subsubsection{Calibration and Removal of Alignment-induced Distortions}
As shown in Figure \ref{fig:telalign}, the distortion variations induced by alignment errors are of low order, and are therefore well sampled by the motion of the diffraction spikes they produce. To compute a 2D map of angular distortions from images of the spikes acquired at different epochs, the data analysis proposed in this paper performs a geometrical interpolation between the spikes. This approach is well suited when no a priori information is available about the nature of the distortions, and is used in the numerical simulation. 

Astrometric distortions induced by alignment errors consist of a limited number of modes (shown in Figure \ref{fig:telalign}), and are likely to be the dominant source of astrometric distortion variation. Table \ref{tab:telalign} shows that a 10 $\mu$m motion of M2 or M3 produces a 1/100th pixel variation in astrometric distortion, which is approximately 10$\times$ larger than the dynamic distortion anticipated from changes of the mirrors' optical figure and detector deformations (image (a) in Figure \ref{fig:numsimexample}) combined. 

A more appropriate approach is to first fit and subtract the known alignment-induced distortion modes from the measured diffraction spikes motion, and then perform the 2D interpolation on the residual spikes motion. There are 16 alignment modes: with the PM serving as a reference, there are 5 modes for M2 and M3 each (3 translations and 2 rotations; the rotation around the conic surface vertex is not included due to rotational symmetry of the surfaces), and 6 modes for the detector plane. Since this is a small number of modes to fit compared to the large number of measurements (number of pixels illuminated by the spikes), this modal subtraction should not otherwise affect the measurement.

\subsubsection{Displacement of the diffraction spikes on the focal plane detector array}
We have assumed, as discussed in Section \ref{ssec:overview}, that the diffraction spikes fall on the same pixels for all observations, and that uncalibrated static detector inter- and intra-pixel sensitivity errors do not contribute to errors in the spikes position measurement. As noted in Section \ref{ssec:overview}, if the spikes fall on different pixels between observations, we may assume that percent-level errors in the detector flat field at high spatial frequency will produce a pixelation centroid error of approximately 1/100th of a pixel for each $\lambda/D$-long segment of a spike, and the overall resulting astrometric error would be at about 0.1 $\mu$as thanks to the averaging of many such segments. {\bf For this error term to be well below 0.1 $\mu$as, as assumed in our error budget, the spikes have to fall on the same pixels for all observations.}

Table \ref{tab:telalign} shows that the positions of optical elements need to be stable at the 100 $\mu$m level (translation) / 10 arcsec level (rotation) for the spikes to stay on the same pixel location on the detector between observations, at the 1/10th of a pixel level. With the tolerances given in the table, a quadratic sum of the distortions would produce a 0.35 pixel rms displacement of the spikes on the focal plane array. This amount of displacement is sufficiently small to allow the same pixels to be used for the spikes location measurement between observations, therefore removing the effect of unknown pixel-to-pixel sensitivity variations on the spikes position measurement accuracy --- as assumed in our error budget. Unknown intra-pixel sensitivity variations will however still impact the measurement with the 0.35 pixel rms spike motion, but this second-order effect will be significantly smaller than the first-order pixel-to-pixel flat-field errors, and should therefore not contribute to astrometric errors at the 0.1$\mu$as level.

\section{Conclusion}

The diffractive pupil telescope concept described in this paper is an attractive solution to perform, in a single mission, simultaneous coronagraphic imaging and astrometric mass determination of exoplanets around nearby stars. The diffractive PM can be designed to allow unperturbed coronagraphic imaging of the central field and wide-field imaging for other astrophysical investigations. The concept is especially attractive if a coronagraphic instrument is installed on a wide-field imaging telescope, as the change required to perform astrometric measurement (addition of dots on the PM) is of moderate cost and will not strongly impact other science goals. Our preliminary analysis indicates that with a 0.3 deg$^2$ ($=$ 0.6 deg diameter) FOV camera on a 1.4 m telescope, a 0.2 $\mu$as astrometric accuracy can be reached in a 2 day observation. Performance is expected to improve rapidly with telescope size, and could be further enhanced with additional metrology hardware, environment control and software removal of instrumental errors identified by careful analysis of data acquired through the mission duration.

The baseline concept adopted in this paper assumes no modification to the coronagraphic wide-field telescope design other than the addition of a regular grid of small dots on the PM of the telescope to create the diffraction spikes used for astrometric calibration. Other aspects of the telescope and instrument design are driven by the need for high contrast coronagraphy and wide-field imaging at the telescope's diffraction limit.

Hardware modifications and calibration of critical components (optics and detector) would further improve astrometric performance, and would increase the robustness of the technique by providing valuable quantitative information on the astrometric distortions and how they vary in time. As shown in Table \ref{tab:errorbudg}, the largest sources of errors, excluding photon noise, are optical distortions (both static and dynamic) and variations in detector geometry between observations. Together, these error terms are responsible for 88.5\% (in quadrature) of the total error budget excluding photon noise. Accurate measurement of the optical surfaces after manufacturing, as well as on-orbit metrology to record the relative position and orientations of the optical elements, could be performed to calibrate part of the astrometric distortions introduced by optics. Calibration of the focal plane array geometry and intra-pixel sensitivity variations can be done prior to launch or on-orbit. \cite{2011RSPSA.467.3550Z} have proposed using interferometric fringes projected on the detector to accurately calibrate the detector, and show that micropixel-level centroiding accuracy is possible, even in the presence of unknown PSF variations. A flat fold mirror between the last powered mirror of the system and the detector array could also be used to dither the wide-field image on the detector, and therefore allows calibration and/or averaging down of error terms due to the detector flat field and detector geometry (N12, N13, N14, N16, and N17), which are responsible for 42.2\% (in quadrature) of the total error budget excluding photon noise. In the optical design shown in Figure \ref{fig:optprinciple}, this mirror could be placed in the pupil plane located between M3 and the detector to avoid introducing additional astrometric distortions (there is no beam walk in a pupil plane).

Simultaneous high-precision astrometry and coronagraphic imaging is a powerful combination for the identification and characterization of exoplanets, with a total scientific return exceeding the sum of what can be derived separately from the two individual measurements. This science gain will be explored and quantified in a future publication.

We are currently developing a laboratory test bed of the concept to validate and refine, when possible, the error budget presented in this paper. Data reduction algorithms will be developed and tested, and laboratory data will increase our ability to link telescope stability and optical quality to astrometric performance. Accurate estimation of the astrometric accuracy will also require detailed modeling of the spacecraft and telescope stability over long timescales. A key goal of these activities is to determine if sub-$\mu$as measurement accuracy can be reached with a conventional wide-field telescope, or if continuous calibration of detector distortions with laser metrology \cite{2011RSPSA.467.3550Z} and active alignment control are required.

We note that a different approach to high-precision astrometric imaging, proposed by \cite{2011ExA...tmp...87M}, is to use a single mirror telescope (no secondary mirror) free of astrometric distortion. While this approach does not require the diffractive pupil calibration described in this paper, it requires a long focal length telescope (at least f/20 for a 1 m diameter aperture) and laser metrology (which may also improve performance for the diffractive pupil concept presented in this paper). The single mirror astrometric telescope concept would use a small number of movable detectors, and does not offer deep wide-field imaging capability for non-exoplanet science. Both concepts are compatible with a coronagraph instrument, although the single mirror astrometric telescope concept would require the coronagraphic imaging to be non-simultaneous with astrometric measurement, unless the light from the central star is shared with a beam splitter.

The diffractive pupil telescope may also enable high-precision astrometry with ground-based telescopes, as it can calibrate instrumental sources of astrometric errors, and atmosphere-induced distortions should average down with exposure time.

\acknowledgments
This work is funded by the NASA Astronomy and Physics Research and Analysis (APRA) program and the State of Arizona Technology Research Initiative Fund (TRIF).

\bibliography{ms}

\appendix

\section{Fundamental Limitations of diffraction-limited imaging astrometry}

In this appendix, the fundamental (no distortions, perfect noiseless detector) limits of astrometric accuracy are reviewed. Results shown in this appendix are used in the numerical model described in Figure \ref{fig:numsimulchart} to account for photon noise, PSF chromaticity, and finite detector sampling.

\subsection{Astrometric accuracy at the diffraction limit}

With an ideal telescope (no wavefront aberration, no central obstruction), the accuracy with which a monochromatic PSF (exact Airy function) can be localized in 2D is, in the photon noise regime.
\begin{equation}
\label{equ:sigma1D}
\sigma_{2D} [\lambda/D] = \sqrt{2}/(\pi \sqrt{N_{ph}})  [\lambda/D] = 0.450/\sqrt{N_{ph}} [\lambda/D] 
\end{equation}
where brackets indicate the unit, $D$ is the telescope diameter, $\lambda$ is the wavelength, and $N_{ph}$ is the total number of photon available for the measurement.
If the PSF position is measured along one axis only, the measurement error is
\begin{equation}
\label{equ:sigma2D}
\sigma_{1D} [\lambda/D] = 1/(\pi \sqrt{N_{ph}}) [\lambda/D]  = 0.318/\sqrt{N_{ph}} [\lambda/D] 
\end{equation}

These measurement accuracies are obtained with an optimal matched filter which optimally weights each pixel according to SNR. Numerically these equations can be verified by starting from the PSF (noted PSF($x$,$y$)) and computing
\begin{equation}
\label{equ:appAsigx}
(1/\sigma_x)^2 = \sum_{x,y} (1/\sigma_x(x,y))^2  \sum_{x,y} \left( \left(\frac{\delta PSF(x,y)}{\delta x}\right)^2 / PSF(x,y) \right)
\end{equation}

with $\sigma_{1D} = \sigma_x = \sigma_y$, and $\sigma_{2D} = \sqrt{2} \times \sigma_{1D}$. Equation (\ref{equ:appAsigx}) shows for each pixel the signal (equal to $\delta$ PSF($x$,$y$)/$\delta x$) and the noise (equal to $\sqrt{PSF(x,y)}$ in the photon noise regime considered here).

\subsection{PSF chromaticity and photon noise}

\begin{figure}
\includegraphics[scale=0.55]{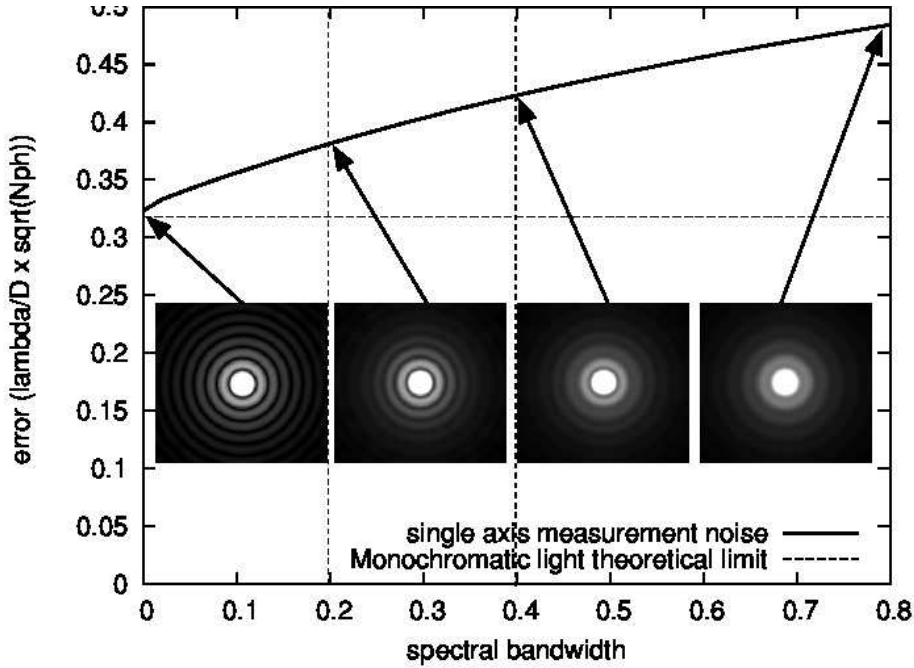} 
\caption{\label{fig:astromaccdl} 
Loss of photon noise limited astrometric measurement accuracy as the spectral bandwidth increases. The data were computed at 10$\times$ Nyquist at the central wavelength, with Equation (A3) used to compute the astrometric error due to photon noise. Polychromatic PSFs are computed as an incoherent sum of 50 monochromatic PSFs, each scaled to the same physical plate scale. In each case, the spectra is flat in photon per unit of $d\lambda$, and extends from $\lambda_0/(1+x/2)$ to $\lambda_0 \times (1+x/2)$, where $x$ is the spectral bandwidth. The sampling is 10$\times$ Nyquist at the central wavelength. 
}
\end{figure}

Equations (\ref{equ:sigma1D}) and (\ref{equ:sigma2D}) assume a monochromatic PSF, but a real polychromatic PSF is not as sharp and will lead to a lower astrometric accuracy for the same total number of photon (as shown in Figure \ref{fig:astromaccdl}). This performance loss is however modest, especially compared to the sensitivity gain offered by including more flux as the spectral bandwidth increases. Figure \ref{fig:astromaccdl} shows that with a 50\% spectral bandwidth, the astrometric error is 38\% larger than it would be in monochromatic light with the same number of photon.

\subsection{Detector Sampling}

\begin{figure}
\includegraphics[scale=0.75]{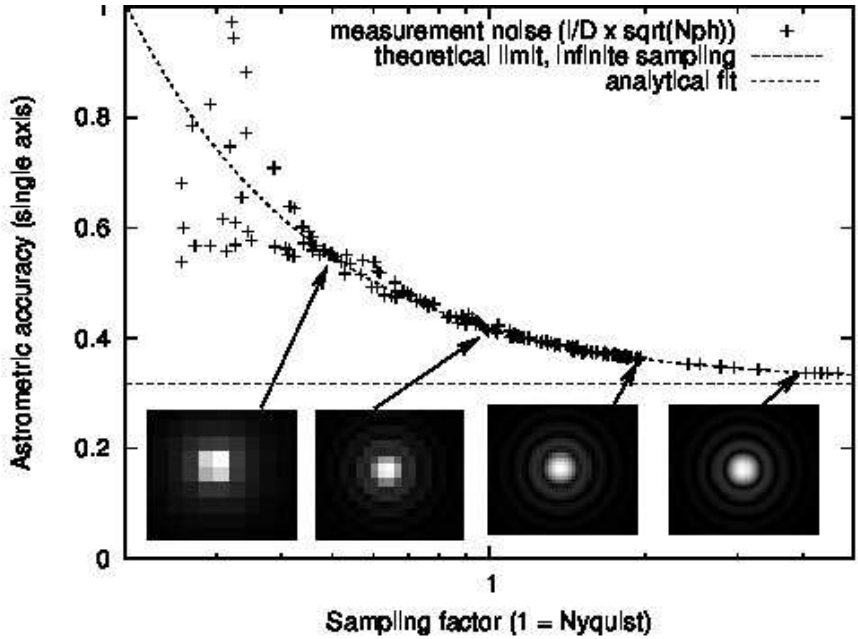} 
\caption{\label{fig:astromaccsampl} 
Single axis photon noise limited astrometric measurement accuracy as a function of detector sampling. Monochromatic simulation. At small sampling, the points fan out between a lower boundary corresponding to the "quad cell" case where the PSF center falls at the corner of 4 pixels (and the astrometric error becomes independent of sampling, as shown by the flatness of this lower boundary), and an upper boundary where the PSF is centered on a pixel.}
\end{figure}

The astrometric error induced by photon noise is well described by the following analytical fit, derived empirically from the points in Figure \ref{fig:astromaccsampl}, and also shown in the figure.
\begin{equation}
\sigma_{1D} [\lambda/D] = N_{ph}^{1/2} \times (1/\pi + 0.1 \times SamplingFactor^{-1.2})  [\lambda/D],
\end{equation}
where $SamplingFactor$ is 0.5 times the linear number of pixel per $\lambda/D$, and is therefore equal to 1 for Nyquist sampling ($SamplingFactor$ = 1.5 means 3 pixel per $\lambda/D$). Figure \ref{fig:astromaccsampl} shows that astrometric accuracy in the photon noise regime continues to improve as the sampling increases beyond the Nyquist limit. Provided that a good model of the PSF exists, the astrometric accuracy remains quite good below Nyquist sampling: at 0.4$\times$ Nyquist (pixel linear size $> \lambda/D$), the 2D astrometric accuracy is still below $1/\sqrt{N_{ph}} \lambda/D$. Below 0.4 Nyquist, the astrometric accuracy is a function of the exact PSF location, with optimal sensitivity when the PSF falls on the corner between pixels.

\begin{figure}
\includegraphics[scale=0.75]{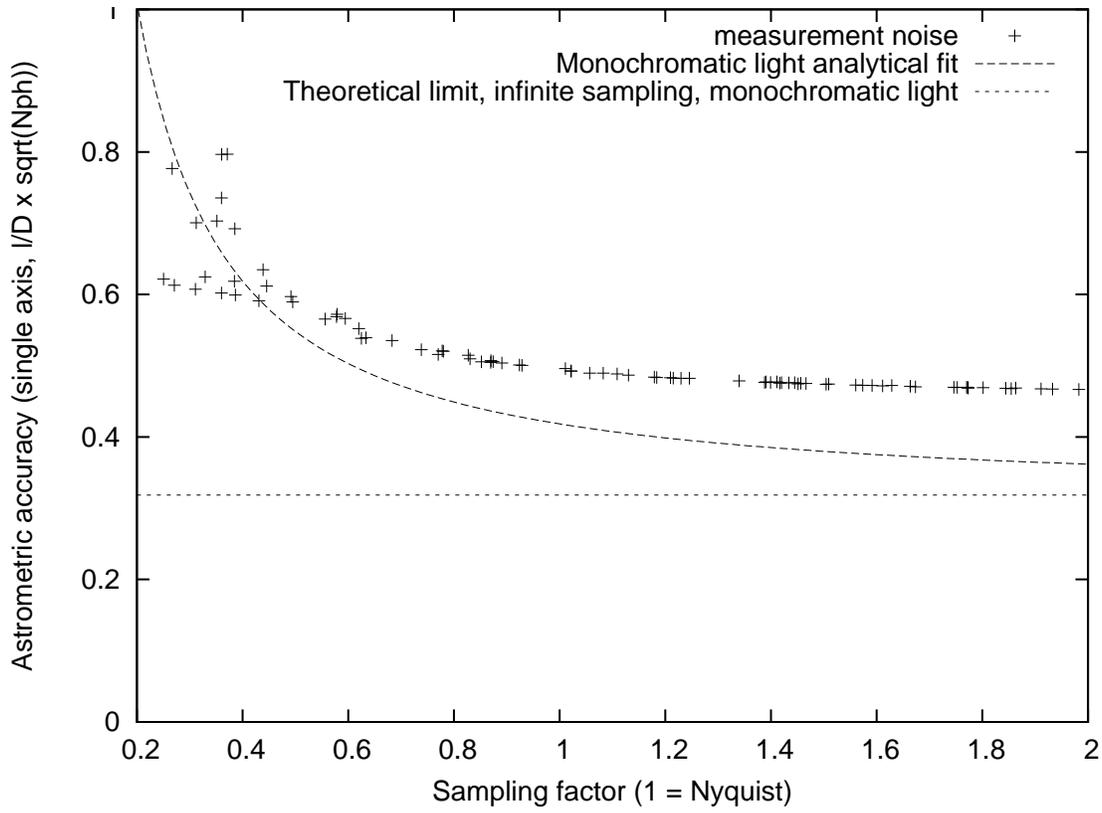} 
\caption{\label{fig:astromaccfinal} 
Single axis photon noise limited astrometric measurement accuracy as a function of detector sampling for a polychromatic measurement. There is little gain beyond Nyquist sampling. }
\end{figure}

The combined effect of finite detector sampling and polychromaticity on astrometric accuracy is shown in Figure \ref{fig:astromaccfinal} in the absence of a background (no zodiacal light). For this figure, the flux is assumed to be constant in units of photon per wavelength bin from 0.52 to 0.78 $\mu$m, and decreases continuously outside this band, with the 50\% level at 0.47 and 0.83 $\mu$m and the 0\% level at 0.4 and 0.9 $\mu$m. For the sampling and astrometric accuracy units, $\lambda = 0.6 \mu$m is assumed.

\end{document}